\documentclass[12pt]{article}

\usepackage{latexsym} 
\usepackage{amssymb} 
\usepackage{amsfonts} 
\usepackage{graphicx} 
\pdfoutput=1

\catcode`\@=11
\def\numberbysection{\@addtoreset{equation}{section}
        \def\theequation{\thesection.\arabic{equation}}}

\def\be{\begin{equation}} 
\def\ee{\end{equation}} 
\def\ba{\begin{eqnarray}} 
\def\ea{\end{eqnarray}}

\def\ov{\overline} 
\def\I{{\rm Im}} 
\def\R{{\rm Re}} 
\def\Z{\mathbb{Z}} 
\def\C{\mathbb{C}} 
\def\RR{\mathbb{R}} 
\def\T{\mathbb{T}} 
\def\nl{\nonumber \\}

\def\wh{\widehat}

\def\a{\alpha} 
\def\b{\beta} 
 
\def\G{\Gamma} 
\def\D{\Delta} 
\def\d{\delta} 
 
\def\eps{\varepsilon} 
\def\z{\zeta}

\def\l{\lambda} 
\def\L{\Lambda}

\def\s{\sigma} 
 
\def\t{\tau} 
\def\f{\phi}

\def\c{\chi} 
\def\w{\omega} 
 
\def\Th{\Theta} 

\def\winf{W_{1+\infty}} 
\def\u1{\widehat{U(1)}}
\def\su2{\widehat{SU(2)}_1}
\def\suem{\widehat{SU(m)}_1}

\def\disp{\displaystyle}

\numberbysection

\begin{document} 
\begin{titlepage}
\begin{center}
\hfill  \quad DFF 450/07/2009 \\

\vskip .3 in
{\LARGE Chiral Partition Functions}

{\LARGE of Quantum Hall Droplets}

\vskip 0.2in
Andrea CAPPELLI \\
{\em INFN \\ Via G. Sansone 1, 50019 Sesto Fiorentino - Firenze, Italy}
\\
Giovanni VIOLA \\
{\em INFN and Dipartimento di Fisica \\ 
Via G. Sansone 1, 50019 Sesto Fiorentino - Firenze, Italy}
\\
\vskip 0.1in
Guillermo~R.~ZEMBA \\
{\em Facultad de Ciencias Fisicomatem\'aticas e Ingenier\'{\i}a, UCA 
\\Av. A. Moreau de Justo 1500,
(1107) Buenos Aires, Argentina\\
and Departamento de F\'{\i}sica, CNEA \\
Av.Libertador 8250, (1429) Buenos Aires, Argentina}
\end{center}
\vskip .1 in
\begin{abstract}

Chiral partition functions of conformal field theory describe the edge
excitations of isolated Hall droplets. They are characterized by an index
specifying the quasiparticle sector and transform among themselves
by a finite-dimensional representation of the modular group.
The partition functions are derived and used to describe electron
transitions leading to Coulomb blockade conductance peaks. 
We find the peak patterns for Abelian hierarchical 
states and non-Abelian Read-Rezayi states, and compare them.
Experimental observation of these features can check the qualitative
properties of the conformal field theory description, such as the
decomposition of the Hilbert space into sectors, involving charged and
neutral parts, and the fusion rules.
\end{abstract}

\vfill

\end{titlepage}
\pagenumbering{arabic}


\section{Introduction}


\subsection{Disk partition functions}

Within the conformal field theory description of edge excitations
\cite{wen}\cite{stern-rev}, the advanced methods of rational conformal
field theories (RCFT) \cite{cft} have been recently applied to discuss
the interference of non-Abelian anyons \cite{na-interf}\cite{tqc} and
to determine the bipartite entanglement entropy of topological ordered
ground states \cite{tee}.
The RCFT methods, and the corresponding properties of topological Chern-Simons
gauge theories, were originally investigated by Verlinde 
\cite{verl}, Witten \cite{jones}, Moore and Seiberg and others \cite{mose}.
These authors analyzed the equations of crossing symmetry (duality) of
$n$-point correlators and the identities relating these functions 
among themselves on general Riemann surfaces.
The fundamental sets of relations were shown to be: 
i) the modular invariance of the partition functions ($1$-point functions)
on the toroidal geometry; 
ii) the crossing symmetry of the $4$-point function on the sphere. 
All other equations follow by ``sewing'' the surfaces with handles
and punctured spheres \cite{mose}.

The rational theories are characterized by a finite set of quasiparticle
excitations with rational values of spin and statistics (scaling dimensions).
The prominent theories of the quantum Hall effect are indeed RCFTs: the
Abelian states, which are (multicomponent) Luttinger liquids described by
charge lattices \cite{hiera}, and the non-Abelian Read-Rezayi states \cite{rr},
that involve $\Z_k$ parafermions \cite{z-k}.  Other theories, such as the
$\winf$ models \cite{w-min} and the Fradkin-Lopez theory \cite{lf} for the
Jain states \cite{jain}, are not rational theories but projections of them,
such that their properties can be traced back to those of RCFTs.  
Therefore, the rational CFTs are of general physical interest.

In the recent literature, the RCFT methods have been reconsidered and
extended \cite{freedman}\cite{tqc}: the duality equations for four and
higher-point functions have been thoroughfully analyzed,
in particular for studying interferometry of non-Abelian anyons
\cite{na-interf}.  On the other hand,  the modular invariant
partition functions have not been extensively discussed.

In this paper, we provide a rather complete and selfcontained discussion of
the partition functions in the QHE setting.  Based on a previous study of
partition functions on the annulus geometry \cite{cz}, we obtain the chiral
partition functions that pertain to the disk geometry and study their
properties \cite{cgz}.  These partition functions describe the excitations of
isolated droplets of Hall fluid and provide a complete definition of their
Hilbert space and its decomposition into sectors; moreover, the fusion rules,
the selection rules for the composition of excitations, are built in.

The general form of the annulus partition function is:
\be
Z_{\rm annulus}=\sum_{\lambda=1}^N\ \theta_\lambda\ \ov\theta_{\lambda} \ ,
\label{z-ann}
\ee
where the index $\l$ runs over the sectors of the theory,
described by the functions $\theta_\l$, that are analytic (resp. anti-analytic)
for the inner (resp. outer) edge.

The annulus partition function is defined on the spacetime torus made by the
edge circle and the compact Euclidean time: it is invariant
under modular transformations, the discrete coordinate reparametrizations that
respect the double periodicity of the torus geometry \cite{cft}.
In RCFTs, modular invariance  is achieved as
follows: the generalized theta functions $\theta_\l$ transform by a
unitary linear representation of dimension $N$, that leaves the
sesquilinear form (\ref{z-ann}) invariant.
An interesting feature is that the number of sectors $N$ is equal to Wen's
topological order \cite{wen}, the degeneracy of Hall ground states on
the compact toroidal space (a general proof of this result is reported
in section 2.2 of \cite{cz}).  Another fact is that the states in each
$\l$ sector form a representation of the maximal chiral algebra of the
RCFT, whose character is $\theta_\l$.

In section 2, we introduce the annulus partition functions in the simpler case
of Laughlin plateaux, with filling fraction $\nu=1/p$ ($p$ odd); we describe
the conditions of modular invariance and solve them to obtain the form
(\ref{z-ann}) with $N=p$.  We show that the geometrical modular conditions
have clear physical meanings in the QHE setting and provide useful building
principia.  In the Laughlin case, the characters $\theta_\l$ resum all
excitations with same fractional charge part, $Q=\l/p +\Z$: 
the corresponding chiral
algebra is the extension of $\u1$ current algebra (Luttinger theory) by a
field of scaling dimension $h=p/2$; the fusion rules for the corresponding
excitations are given by the additive group $\Z_p$.

In Section 2.3, the disk partition functions are obtained by 
taking the limit of $Z_{\rm annulus}$ when the inner circle 
shrinks to zero, leading to:
\be 
Z_{\rm disk}^{(\a)}=\theta_\a\ ,
\label{z-disk1} 
\ee 
where the sector $\l=\a$ is chosen according to the type of 
quasiparticles in the bulk \cite{cgz}\cite{schou}.
Besides the Laughlin case, in this paper we provide the expressions
of $Z_{\rm disk}$ for 
general multicomponent Luttinger liquid theories, including
the Jain states, and for the non-Abelian Read-Rezayi states,
using new and known results of the corresponding
annulus partition functions \cite{cz}\cite{cgt2}.

\subsection{Coulomb blockade in quantum Hall droplets}

\begin{figure}[t]
\begin{center}
\includegraphics[width=6cm]{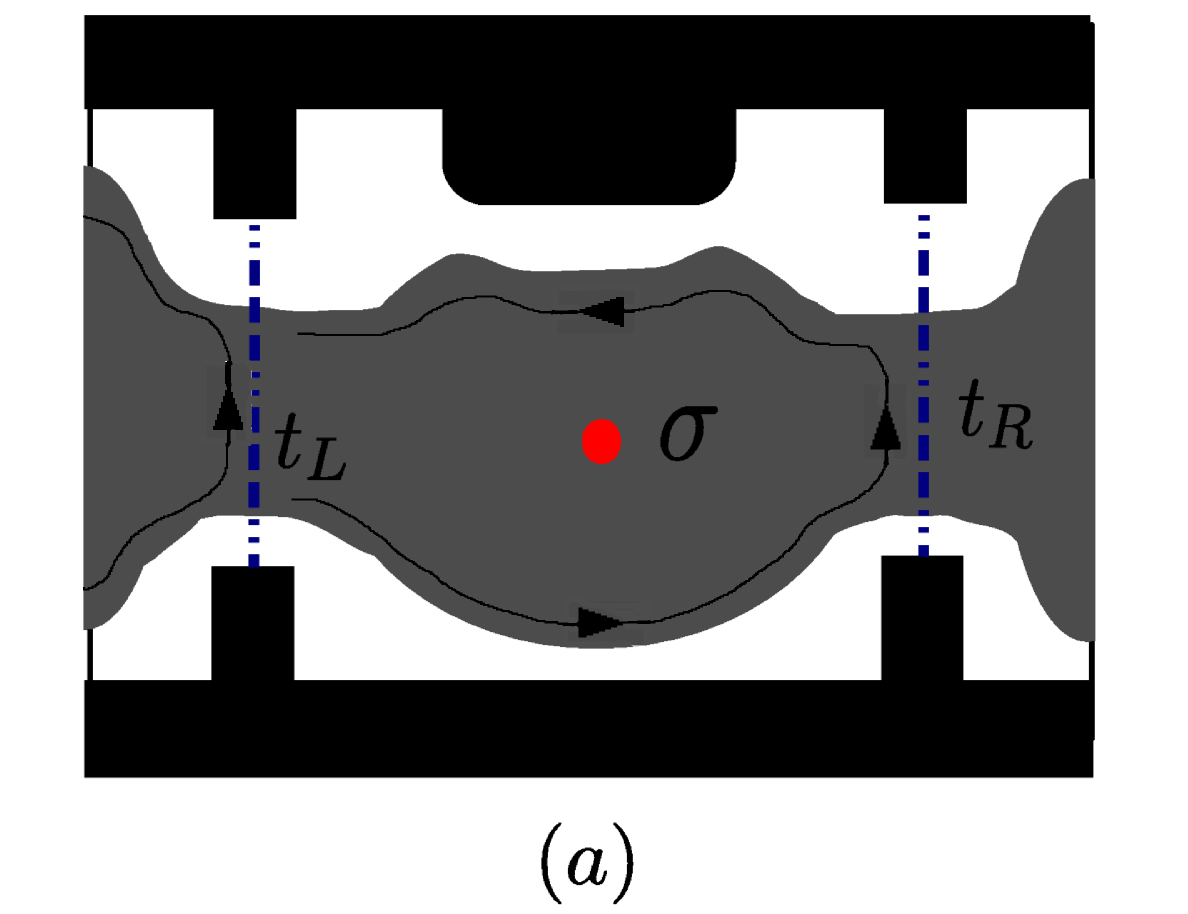}
\hspace{1cm}
\includegraphics[width=6cm]{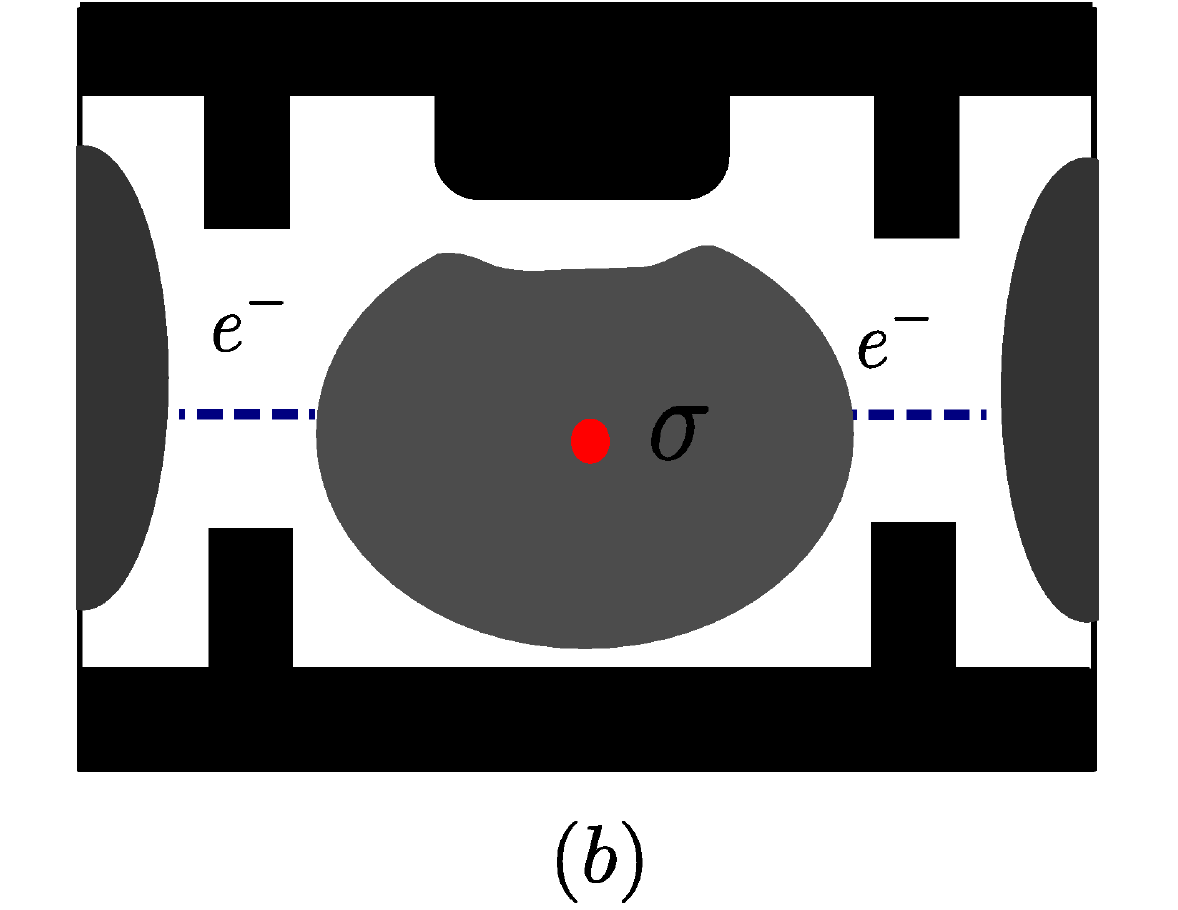}
\end{center}
\caption{The quantum Hall interferometer: the electron fluid is drawn in the
(a) weak and (b) strong backscattering limits.}
\label{fig-dev}
\end{figure}

In Fig. \ref{fig-dev} is drawn the geometry of the quantum Hall 
interferometer, namely a bar-shaped sample with two constrictions \cite{interf}.
In this device, one can consider two opposite regimes of weak
and strong backscattering of edge excitations at the constrictions:
the interference of edge waves is best achieved in the weak backscattering 
limit $(a)$, while the Coulomb blockade takes place
for strong backscattering $(b)$.
In the latter limit, an isolated droplet of Hall fluid is formed and
only electrons can tunnel into the droplet. The electric potential
difference is counterbalanced by the electrostatic charging energy
leading to conductance peaks at exact matching. One can observe
characteristic peak patterns upon varying: i) the area of the dot by
means of a side modulation gate or ii) by tuning the magnetic field (a
third possibility not discussed here could be charging 
an antidot engineered inside the droplet).
The Coulomb blockade in quantum Hall droplets has been
considered in \cite{halp}\cite{stern}\cite{schou}, where it was shown to
provide interesting tests of the conformal field theory description. 
Indications of experimental feasibility have 
been recently reported \cite{cb-exp}.

In this paper, we obtain the peak patterns from the knowledge of the disk
partition function (\ref{z-disk1}) \cite{cgz}.  Each $Z_{\rm disk}^{(\a)}$
resums all excitations corresponding to adding electrons to the droplet within
the $\a$ sector (of given fractional charge): 
level deformations and degeneracies are all accounted there.
We study the peak patterns in the $(S,B)$ plane, corresponding to
simultaneous changes of area and magnetic field, and discuss the bulk-edge
relaxation (recombination) of neutral excitations.

In section 3, we give a detailed account of the Coulomb blockade
in the Jain hierarchical states already presented in \cite{cgz},
as well as the results in two alternative theories for the same states.
In section 4, we discuss the case of Read-Rezayi non-Abelian states, 
extending the results \cite{halp}\cite{stern}\cite{schou}.
The peaks follow a periodic pattern with 
a modulation in the separations that is due to the presence of
neutral excitations. Although the peak patterns are
qualitatively similar in the Abelian and non-Abelian cases,
there are specific differences:

i) in the Abelian Jain hierarchical case, the energy levels possess
 multiplicities characteristic of the multicomponent 
fluids, that can be observed experimentally in the peak patterns;
moreover, all features are independent of the bulk quasiparticle 
sector $\a$, and bulk-edge relaxations are not possible.

ii) in the non-Abelian Read-Rezayi states, there are no degeneracies
and the peak patterns depend on the sector; furthermore, bulk-edge
relaxations could be possible, and would wipe out the dependence on the
sector.

In summary, the patterns of Coulomb blockade peaks can test
the qualitative properties of the Hilbert space of
conformal field theory: the fusion rules, the sectors and
the multiplicities of excitations. These features are
manifestly shown by the disk partition functions.


\section{Partition functions in QHE}

\subsection{Annulus geometry}

\begin{figure}[t]
\begin{center}
\includegraphics[width=6cm]{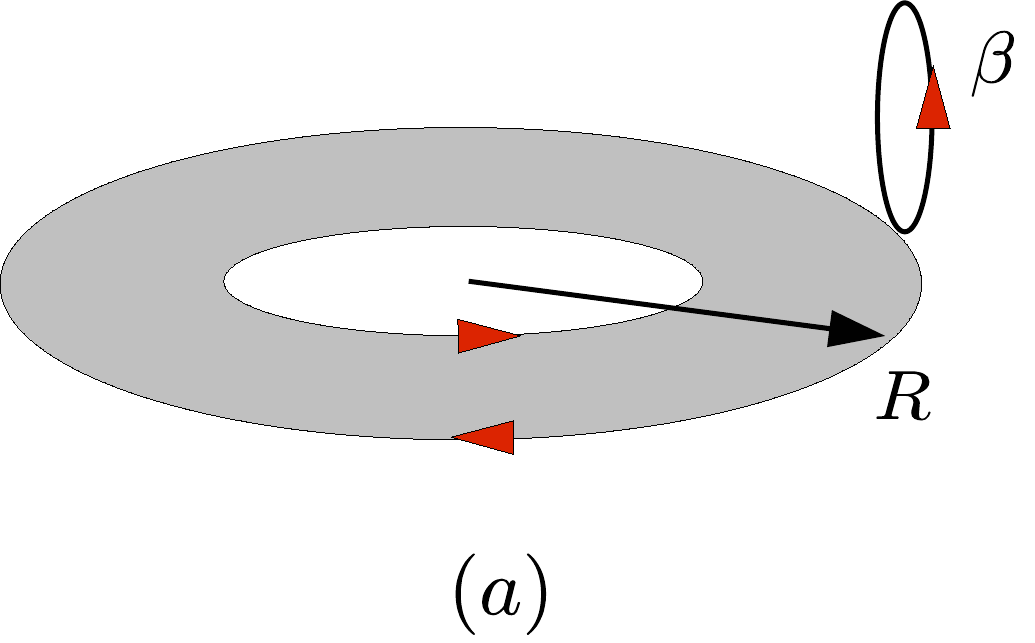}
\hspace{1cm}
\includegraphics[width=6cm]{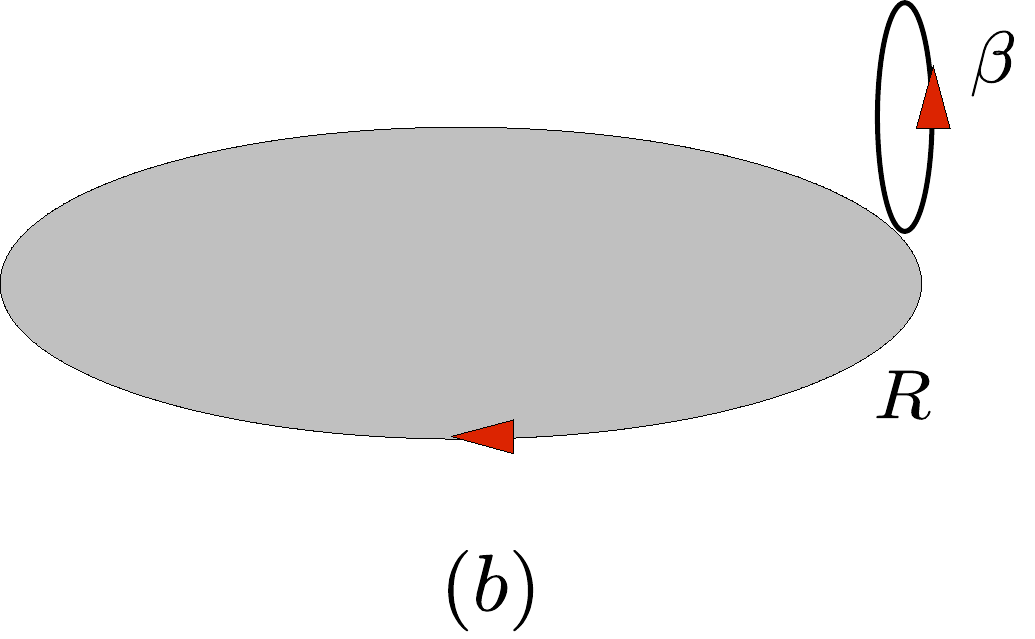}
\end{center}
\caption{The (a) annulus and (b) disk geometries.}
\label{z-geom}
\end{figure}

The Hilbert space of a RCFT is made by a finite number of representations of
the maximally extended (chiral) symmetry algebra, which contains the Virasoro
conformal algebra as a subalgebra \cite{cft}.  The representations are encoded
in the partition function of the Euclidean theory defined on the geometry of a
space-time torus $S^1\times S^1$.
Moreover, any RCFT is associated to a Chern-Simons theory, and the
torus partition function in the former theory corresponds to a
path-integral amplitude for the latter theory on the manifold
$S^1\times S^1\times {\RR}$, where ${\RR}$ is the time axis
\cite{jones}\cite{c-s}.

In the quantum Hall effect, we can consider a spatial annulus with Euclidean
compact time of period $\b$, the inverse temperature: the topology of this
space-time manifold is ${\cal M}=S^1\times S^1\times I$, where $I$ is the
finite interval of the radial coordinate.  The partition function of edge
excitations is defined on the boundary $\partial{\cal M}$, corresponding to
two copies of a space-time torus (see Fig. \ref{z-geom}(a)).  The excitations
are chiral and anti-chiral waves on the outer $(R)$ and inner $(L)$ edges,
respectively.
The annulus coordinates are $(\varphi, t_E,r)$, with
$r\in {\rm I}=[R_L,R_R]$ and $t_E\equiv t_E+\beta$, 
$\varphi\equiv \varphi+2\pi$.
We assume that there are no static bulk excitation inside the annulus;
they could be added afterwards but will not be necessary.

We illustrate the method and set the notation by first repeating  
the analysis of the simpler Laughlin states \cite{cz}.
The following spectrum was obtained for the excitations on each edge
from the canonical quantization of the chiral Luttinger liquid
\cite{cdtz}:
\be
\nu={1\over p}\ , \quad Q={n\over p}\ , \quad L_0={n^2\over 2p}\ ,
\qquad n \in \Z\ , \ p=1,3,5,\dots\ .
\label{specu1}\ee
Each pair of values $(Q,L_0)$ are weights of a (highest-weight)
representation of the $\u1$ affine (current) algebra of the conformal
theory with central charge $c=1$, 
$L_0$ being the zero mode of the Virasoro algebra.  

We shall obtain the partition functions $Z_{\rm annulus}$ by using these
data and by imposing the modular invariance conditions.
We start by the definition \cite{cft}: 
\be
{\rm Z}_{\rm annulus}\left(\tau,\zeta \right)\ = \ {\cal K}
\ {\rm Tr}\ \left[ {\rm e}^{i2\pi \left(
\tau\left(L_0^L- c/24\right) - \ov\tau\left(L_0^R- c/24\right) +
\zeta Q^L+ \ov\zeta Q^R \right) } \right]\ ,
\label{zdef}\ee
where the trace is over the states of the Hilbert space, ${\cal K}$
is a normalization to be described later and $(\tau,\zeta)$ are complex 
numbers. We recognize (\ref{zdef}) as the grand-canonical partition function:
the total Hamiltonian and spin of excitations are:
\be
H={v_R\over R}\left(L^R_0- {c\over 24}\right) +
{v_L\over R}\left(L^L_0- {c\over 24}\right)
+V_o\left(Q^L-Q^R\right) +\ {\rm const.},
\qquad J=L^L_0-L^R_0\ .
\label{hconf}\ee
The energy is proportional to the conformal dimension, $E =(v/R)(L_0 -c/24)$,
with $c$ parameterizing the Casimir energy \cite{cft}. 
The real and imaginary parts of $(\tau, \zeta)$ are related to
the inverse temperature $\beta$ and Fermi velocity $v$, 
$2\pi R\ \I\tau= v \beta>0$, 
the ``torsion'' $\R\tau$, the chemical potential
$\mu\b=2\pi \R\zeta$ and the electric potential between the two
edges $V_o\b=2\pi \I\zeta$.
Since the Virasoro dimension is roughly the square of the charge,
the partition sum is convergent for $\I\tau >0$ and
$\zeta\in {\C}$.
Let us momentarily choose a symmetric Hamiltonian for the two edges
by adjusting the velocities of propagation of excitations,
$v_L/R_L=v_R/R_R$. 

As usual in RCFT, one can divide the trace in (\ref{zdef}) into a sum over
pairs of highest-weight $\u1$ representations (one for each edge) 
and then sum over the states within each representation. 
The latter give rise to the $\u1$ characters \cite{cft}:
\be
Ch\left(Q,L_0\right)\ = \left.{\rm Tr}\right\vert_{\u1}\
\left[ {\rm e}^{i2\pi \left(
\tau\left(L_0- c/24\right) + \zeta Q\right) } \right]\ = \
{q^{L_0}\ w^Q \over \eta(q)}\ ,
\label{charu1}\ee
where $\eta$ is the Dedekind function,
\be
\eta\left( q\right) =q^{1/24}\ \prod_{k=1}^\infty \left( 1-q^k \right)\ ,
\qquad q={\rm e}^{\disp i2\pi \tau}\ ,\quad w={\rm e}^{\disp i2\pi \zeta}\ .
\label{dede0}\ee
Any conformal field theory with $c\ge 1$ contains an infinity of
Virasoro (and $\u1$) representations \cite{cft}; therefore, we must further
regroup the $\u1$ characters into characters $\theta_\l(\t,\z)$ of an extended
algebra in order to get the finite-dimensional decomposition:
\be
Z_{\rm annulus}\ =\ \sum_{\lambda,\ov\lambda=1}^N\
{\cal N}_{\ \l,\mu}\ \theta_\l\
\ov\theta^c_\mu\ .
\label{rcftz}\ee
In this equation, the bar denotes complex conjugation and the suffix
$(c)$ is the charge conjugation $C$, acting by: $ Q \to -Q$, $\theta \to
\theta^c$.  
The inner (resp. outer) excitations are described by
$\theta_\l$ (resp. $\theta_\mu^c$), according to the definition (\ref{zdef}). 
The coefficients ${\cal N}_{\ \l,\mu}$ are positive integers giving the
multiplicities of sectors of excitations: they are not known in general, but
in cases of explicit path-integral calculations.  In the following, we
shall self-consistently determine the ${\cal N}_{\ \l,\mu}$ by imposing
modular invariance and some physical requirements.

\subsection{ Modular invariance conditions}

The torus geometry $\T$ can be described as the quotient of the
complex plane $\C$ by the lattice of translations $\L$ generated by the
two periods $\w_1$ and $\w_2$: 
\be
\T =\frac{\C}{\L}\ ,\qquad
z \sim z +n_1 \w_1 +n_2 \w_2 \ , \qquad z \in \C, \ \ n_1,n_2 \in \Z \ .
\ee
The same lattice can also be generated by another pair of periods
$\omega'_1$ and $\omega'_2$, that 
are related by linear integer transformations with 
unit determinant, i.e by $SL(2,\Z)$ mappings.
Owing to scaling invariance, the torus is characterized by 
the complex modulus, $\tau =\omega_2/\omega_1$, with $\I \t >0$.
Equivalent coordinates on the torus are related by the modular 
transformations,
\be
\tau^{'} = \frac{a\tau +b}{c\tau +d}\ , \qquad a,b,c,d \in \Z\ ,
\qquad ad-bc = 1\ .
\label{fracl}
\ee
The modular group is thus given by $\Gamma \equiv PSL(2,\Z )= SL(2,\Z )/\Z_2$
(the quotient is over the global sign of transformations):
it is known that the group has two generators,
$T\ :\ \tau\to\tau + 1$
and $S\ :\ \tau\to -1/\tau$, satisfying the relations
$S^2=\left(ST\right)^3 =C$, where $C$ is the charge conjugation matrix,
$C^2=1$ \cite{cft} 
(more details are given in Appendix A).

The invariance of the partition function under modular transformations
is therefore given by:
\be
Z_{\rm annulus}\left(\frac{-1}{\t},\frac{-\z}{\t} \right) = 
Z_{\rm annulus}\left(\t+1,\z \right) = Z_{\rm annulus}\left(\t,\z\right)\ .
\label{minv-zed}
\ee
Note that the other parameter $\z$ transforms under (\ref{fracl})
as a coordinate on the torus, that acquires the scale factor: 
$\z'=\z/(c\t+d)$.

In the following discussion, these geometrical conditions will be
obtained from physical requirements on the QHE system, thus showing
that they are appropriate for the CFT description of edge excitations.

The most interesting modular transformation exchanges the two periods
of the torus,
\be
S:\qquad Z\left(-{1\over\tau},-{\zeta\over\tau}\right)=
Z\left(\tau,\zeta\right)\ .
\label{scond}\ee
From the geometrical standpoint, the equivalence of the
periods is apparent in the Chern-Simons theory:
the RCFT partition function corresponds to an amplitude 
on ${\cal M}=S^1\times S^1\times {\rm I}$, where
time is $t_E\in {\rm I}$ and the two spatial periods 
are on the same footing \cite{jones}\cite{c-s}.

From the physical point of view, the $S$ invariance amounts to a completeness 
condition for the spectrum of the RCFT: upon exchanging time and space,
it roughly imposes that the set of states at any time $t_E=t_o$ is 
the same as that ensuring time propagation \cite{cft}.
Furthermore, in RCFT partition functions (\ref{rcftz}), 
the $S$ transformation is realized by a linear transformation of the characters,
\be  
\theta_\a\left(-{1\over\tau},-{\zeta\over\tau}\right)=e^{i\varphi}
\sum_{\b=1}^N\ S_{\a\b}\ \theta_\b\left(\t,\z\right) \ ,
\label{linear-s}
\ee 
($\varphi$ is an overall phase to be specified later).
Thanks to the Verlinde formula \cite{verl}, the matrix $S_{\a\b}$ determines
the fusion rules of the RCFT that express the consistency and completeness of
the operator content of the theory.

Let us remark that the completeness of the operator content can also
be enforced by solving the crossing symmetry of $4$-point functions,
where the fusion rules determine the possible intermediate channels.
However, several examples in the literature show that crossing
symmetry may be satisfied on smaller set of sectors than those required by the
$S$ modular invariance; 
the latter is a stronger consistency condition requiring 
``maximal'' extension of the operator content \cite{cft}.

The invariance under the $T$ transformation has the following
 physical motivation. The partition function should
describe physical excitations of the whole sample that can be measured
in conduction experiments: these are electron-like and have integer or
half-integer spin.  Therefore, anyon excitations should pair on the
two edges to form them.  This condition is enforced by,
\be 
T^2:\qquad
Z\left(\tau +2,\zeta\right) \equiv {\rm Tr}\ 
\left[\cdots {\rm e}^{\ i2\pi\ 2\left(L^L_0 -L^R_0\right)} \right] = 
Z\left(\tau, \zeta\right)\ .
\label{tcond}
\ee
The presence of fermionic states in the QHE implies the weaker modular
invariance $T^2$: actually, $S$ and $T^2$ generate a subgroup of the modular
group, $\G_\theta \subset \G$, as discussed in appendix A.

The partition function should also be invariant under independent
transformations of the coordinate $\z$, that express its double periodicity:
$\z\sim\z+1\sim \z+\t$.  From the physical point of view, these geometrical
conditions correspond to requirements on the charge spectrum.

In absence of bulk quasiparticles, the edge excitations should have global
integer charge, $Q^L+Q^R \in \Z$, which measures the number of electrons
injected into the system by attaching leads to the two edges.
This condition reads,
\be
U:\qquad Z\left(\tau,\zeta +1\right) \equiv
{\rm Tr}\ \left[\cdots {\rm e}^{\ i2\pi\left(Q^R +Q^L\right)}
\right] = Z\left(\tau, \zeta\right)\ .
\label{ucond}\ee
It follows that fractionally charged excitations
at one edge must pair with complementary ones on the other
boundary. Consider, for example, $\nu=1/3$; we can imagine to drop in an
electron, which splits into the pair $(Q^L,Q^R)=(1/3,2/3)$, 
or into the others, $(0,1),(2/3,1/3),(1,0)$. 

The different splittings should all be possible and be equally
accounted by the partition function, leading to a further invariance.
They are related one to another by tuning the electric potential
$V_o$ in (\ref{hconf}): the change 
corresponding to adding one flux quantum inside the annulus is,
 $V_o \to V_o+1/R$ (in our notations $e=c=\hbar=1$), and corresponds to
$\zeta\to\zeta +\tau$.
The invariance of the partition function is therefore, 
\be
V:\qquad Z\left(\tau,\zeta +\tau\right) = Z\left(\tau, \zeta\right)\ .
\label{vcond}\ee
The transformation of the partition function under $\zeta\to\zeta +\tau$
is called ``spectral flow'' \cite{cdtz}\cite{cz}\cite{geo2}, for reasons
that will be clear momentarily.

We now solve the conditions $(T^2,S,U,V)$ for the $c=1$ theory.  First
consider the $U$ condition, $Q^L+Q^R \in \Z$: we collect left $\u1$
representations which have integer spaced charges, $Q^L=\lambda/p+\Z$, and
later combine them with the corresponding right representations.
The sums of $\u1$ characters give 
theta functions with rational characteristics:
\ba
\theta_\lambda &=&
{\rm e}^{\disp -{\pi\over p}{\left(\I\zeta\right)^2\over \I\tau} }
\ {1\over \eta}\ 
K_\l \left(\t,\z ; p\right) 
\nonumber\\
&=& {\rm e}^{\disp -{\pi\over p}{\left(\I\zeta\right)^2\over \I\tau} }
\ {1\over \eta}\ \sum_{k\in \Z}\
{\rm e}^{\disp\ i2\pi \left(\tau{(pk+\lambda)^2\over 2p} +
\zeta\left({\lambda\over p}+k\right)\right) }\ ,
\label{thetaf}\ea
indeed showing $Q^L=\lambda/p + \Z$, $\lambda=1,2,\dots,p$. 
(The non-analytic prefactor is explained later).
The transformations $T^2,S,U,V$ of these generalized characters are
found to be (see Appendix A),
\ba
T^2:\ \theta_\lambda\left(\tau+2,\zeta\right)\ & = &
{\rm e}^{ i2\pi\left({\lambda^2\over p}-{1\over 12} \right) }
\theta_\lambda\left(\tau,\zeta\right)\ ,
\nonumber\\
S:\ \theta_\lambda\left(-{1\over\tau},-{\zeta\over\tau}\right) & = &
{ {\rm e}^{i{\pi\over p}\R {\zeta^2 \over \tau} } \over\sqrt{p}}
\ \sum_{\lambda^\prime=0}^{p-1}\
{\rm e}^{ i2\pi{\lambda\lambda^\prime\over p} } \
\theta_{\lambda^\prime}\left(\tau,\zeta\right)\ ,
\label{chitr}\\
U:\ \theta_\lambda\left(\tau,\zeta+1\right)\ & = &
{\rm e}^{ i2\pi\lambda/ p } \
\theta_\lambda\left(\tau,\zeta\right)\ ,
\nonumber\\
V:\ \theta_\lambda\left(\tau,\zeta+\tau\right)\ & = &
{\rm e}^{ -i{2\pi\over p}\left(\R\zeta +\R{\tau\over 2} \right) }
\theta_{\lambda+1}\left(\tau,\zeta\right)\ .
\nonumber
\ea
These transformations show that the generalized characters $\theta_\l$ carry a
unitary representation of the modular group, which is projective
for $\zeta\neq 0$ (the composition law is verified up to a phase).

The corresponding sums of right $\u1$ representations
are given by $\ov\theta^c_{\mu}$ carrying charge $Q^R=\mu/p+{\Z}$.
Finally, the $U$ condition (\ref{ucond}), applied to $Z_{\rm annulus}$ 
(\ref{rcftz}), requires that left and right charges obey:
$\l+\mu=0 $ mod $p$. This form of the partition function 
also satisfies the other modular conditions, $T^2,S,V$,  by unitarity. 
We finally obtain the modular invariant annulus partition 
function of Laughlin's plateaus ($c=1$):
\be
Z_{\rm annulus}=\sum_{\lambda=1}^p\ \theta_\l\ \ov\theta_\l \ .
\label{zedone}
\ee
The coefficients, ${\cal N}_{\ \l,\mu}= \delta^{(p)}_{\l+\mu,0}$,
specify the multiplicities and pairings of sectors.

Let us add some remarks on this result. 

i) The $U$ condition on the charge spectrum dominates the other ones,
because it is linear in the character index $\l$: it leads to a unique
modular invariant partition function that is left-right diagonal.  In
other applications of RCFTs, as e.g. in statistical mechanics, the
$\z$ variable is not usually considered, as well as the corresponding
$U,V$ conditions. Therefore, the spectrum of solutions of the
remaining conditions, $T,S$, can be rather rich, leading to left-right
non-diagonal invariants, such as the ADE classification of $c<1$
minimal theories \cite{ade}\cite{cft}.  In the QHE setting,
non-diagonal invariants could possible when neutral sectors of
excitations (not constrained by $U$) are included in the theory.  
Examples for multicomponent Luttinger liquids have been
found in \cite{cz}, but their physical relevance is unclear.  In this
paper, we shall only consider diagonal modular invariants.

ii) The $V$ transformation illustrates the spectral flow: the
addition of a quantum of flux through the center of the annulus is a
symmetry of the Hamiltonian but causes a drift of
the quantum states among themselves.
Indeed, the Hamiltonian (\ref{hconf}) was made invariant by adding a constant 
capacitive energy in each sector, $E_c=RV_o^2/2p$, through
the non-analytic prefactor in the characters (\ref{thetaf}),
$\exp\left(-\nu\pi \left(\I\zeta\right)^2/ \I\tau\right) $
\cite{cz}\cite{geo2}.
The modified spectrum of excitations (\ref{specu1}) is:
\be
E_{n_L,n_R} ={v\over R}{1\over 2p}\left[
\left(n_L+RV_o\right)^2\ +\ \left(n_R-RV_o\right)^2 \right]\ ,
\label{tunspec}\ee
that has vanishing minimum in both edges for any value of $V_o$.
The spectral flow, $\z\to\z+\tau$, was actually first
discussed in Laughlin's thought experiment defining the
fractional charge \cite{laugh}\cite{geo2}.
Since the charge transported between the two edges 
by adding a flux quantum is equal to the Hall conductivity, we find
that, $\theta_\lambda(\zeta+\tau)\propto\theta_{\lambda+1}(\zeta)$, 
does correspond to $\nu=1/p$. This provides a
check of the normalization of $\zeta$.

iii) In the expressions (\ref{zedone}) one can also verify that all the
excitations have integer monodromies with respect to the electrons:
\be 
J[n_e]+J[n]-J[n_e+n] \in \Z\ ,
\label{estat}\ee
where $(n_e=p,n)$ are the integers in (\ref{specu1}) corresponding to one 
electron and a generic excitation, respectively.

iv) The generalized characters $\theta_\lambda$ (\ref{thetaf}) correspond
to the extension of the $\u1$ algebra by a chiral current $J_p$. This
field is included in the vacuum representation of the extended
algebra, and can be identified from the first non-trivial term in the
expansion of $\theta_0$ into $\u1$ characters: $J_p$ has half-integer
dimension $L_0=p/2$ and unit charge.
The highest-weight representations of this extended RCFT algebra
correspond to generalized primary fields $\phi_\lambda$, whose charge
is defined modulo one (the charge of $J_p$); namely, they collect all
chiral excitations with the same fractional charge.  The fusion rules
for these fields are clearly given by the addition of charges modulo $p$,
\be
\phi_i \cdot \phi_j \equiv N_{ij}^k \ \phi_k\ , \qquad
N_{ij}^k = \delta^{(p)}_{i+j,k}\ , \quad i,j,k \in \Z_p\ ,
\label{fusion-one}
\ee
that closes on the finite set of $p$ elements.

The same fusion algebra can be obtained by using the Verlinde formula
\cite{verl},
\be
N_{ijk}=\sum_{n=1}^p \frac{S_{in}S_{jn}S_{kn}}{S_{0n}}\ .
\label{verlinde-eq}
\ee
Inserting the $S$-modular transformation obtained in
(\ref{chitr}), we find: $N_{ijk}=\delta^{(p)}_{i+j+k,0} $, that
is equivalent to (\ref{fusion-one}) by lowering one index
with the charge-conjugation matrix $C_{ij}=N_{ij0}$ \cite{cz}.


\subsection{Disk geometry}

From the annulus partition function (\ref{zedone}), we can deduce the
disk partition function by letting the inner radius to vanish, $
R_L\to 0$ (see Fig. \ref{z-geom}(b)).  To this effect, the variable
$\ov{\t}$ in $\ov{\theta}_\l$ should be taken independent of $\t$: $\I \t
\neq - \I \ov\t$, $v_R/R_R\neq v_L/R_L$. The annulus partition
function is no longer a real positive quantity but remains modular
invariant, up to irrelevant global phases.
In the limit $ R_L\to 0$, the $\ov{\theta}_\l$ are dominated by their  
$|q|\to 0$ behavior: therefore, only the 
ground state sector remains in (\ref{zedone}), leading
to $\ov{\theta}_\l\to \d_{\l,0}$, up to zero-point energy contributions. 
We find: $Z_{\rm disk}^{(0)}=\theta_0 $. 
If, however, there are quasiparticles in the bulk with charge, 
$Q_{\rm Bulk}=-\a/q$, the condition of total integer charge selects 
another sector, leading to:
\be
Z_{\rm disk}^{(\a)}=\theta_\a \ .
\label{z-disk} 
\ee 
Therefore, the disk partition functions are given by the chiral generalized
characters $\theta_\a$, whose index is selected by the bulk boundary 
conditions. 
The set of functions is modular covariant, i.e. it carries a
unitary finite-dimensional representation of the modular group, as shown by
(\ref{chitr}).

  These partition functions describe the edge physics of isolated Hall
droplets with static bulk quasiparticles. 
Note that each sector has a specific lower-state energy 
that has been discarded in (\ref{z-disk}): indeed, 
it is difficult to compare edge energies of different sectors 
in the disk geometry, because they depend on
the external work for adding gapful bulk quasiparticles
and other environmental effects \cite{cdtz}.
We also remark that the identification of $ Z_{\rm disk}^{(\a)}$ from 
$Z_{\rm annulus}$ is unique because the sectors relative to the 
chiral excitations on the outer edge of the annulus 
(accounted by $\theta_\l$) are paired to those of the inner edge 
($\ov{\theta}_\l$).


\section{Coulomb blockade in Abelian states}

\subsection{Conductance peaks in Laughlin states} 

The Coulomb blockade takes place when an electron tunnels in a small
quantum dot: the current cannot flow freely because 
the charging energy may overcame the work done by
the electric potential, 
\be
\D E(n)= -neV + \frac{(ne)^2}{2C}\ , \qquad \D Q =-ne \ ,
\label{cb-class}
\ee 
where $C$ is the capacitance and $V$ the potential. 
It follows that tunneling is possible when the two terms compensate
exactly, $\D E(n)=0$, leading to isolated peaks in the current because
the charge is quantized.

In QHE droplet of Fig. \ref{fig-dev},
the corresponding stationary condition for Coulomb blockade peaks is: 
\be
E_{S,B}(n+1) = E_{S,B}(n)\ .
\label{cb-quan}
\ee 
Here, $E_{S,B}(n)$ are the energies for adding $n$ 
electrons to the edge, that depend on external parameters such as
the droplet area $S$ and the magnetic field $B$.

The dependence on area deformations $\D S$ can be included in the 
edge spectrum of Laughlin states (\ref{specu1}) as follows \cite{stern}. 
The variation of the droplet area induces
a deviation of the background charge $Q_{\rm bkg}$ with respect to its
(vanishing) equilibrium value, yielding a contribution to the charge
accumulated at the edge, $Q\to \left(Q -Q_{\rm bkg}\right)$. 
The edge energies acquire a electrostatic contribution 
that can be derived near equilibrium  by observing that,
$E \propto \left(Q -Q_{\rm bkg}\right)^2$. 
For $\nu=1/p$, we obtain: 
\be
E_{\l,\s}(n) = \frac{v}{R} \frac{\left(\l+p n-\s \right)^2}{2p} \ , 
\qquad \s= \frac{ B \D  S}{\phi_o} \ ,
\label{def-en} 
\ee
where $\s$ is a dimensionless measure of area deformations.

\begin{figure}[t]
\begin{center}
\includegraphics[width=14cm]{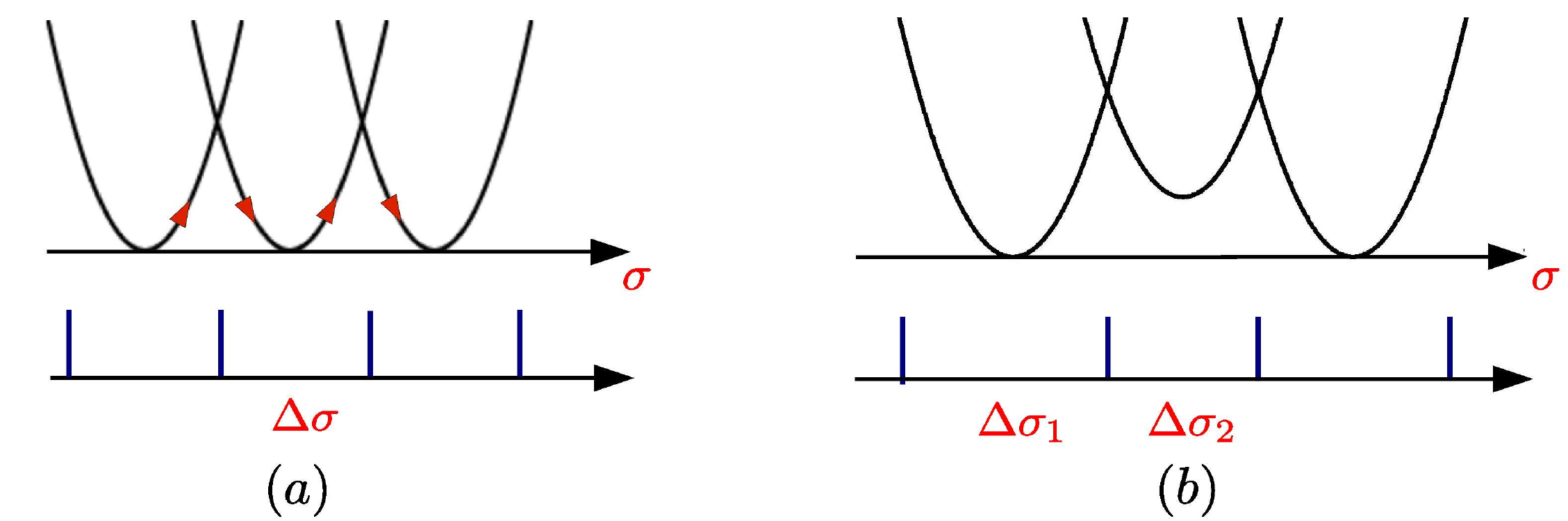}
\end{center}
\caption{Energy levels as a function of the area deformation $\D \s$:
(a) Laughlin states and (b) $m=2$ Jain hierarchical states.}
\label{lev-fig}
\end{figure}

The Coulomb peaks are obtained by looking for degenerate
energy values  with $\D Q =n$, i.e.  within the same
fractional charge sector $\l$: in Fig. \ref{lev-fig}(a), we show the
dependence of $E_{0,\s}(n)$ on $\s$:
the degeneracy condition (\ref{cb-quan}) is satisfied
at the midpoints between two consecutive parabolas;
there, the electrons can tunnel into the dot yielding the
conductance peaks. The separation between two consecutive peaks is:
\be
\D \s =p =\frac{1}{\nu} \ , \qquad \D S = \frac{e}{n_o}\ .
\label{peak1}
\ee 
This result is consistent with the classical picture for which the 
change in the area precisely matches the value required for incorporating
one electron at the average density $n_o$ \cite{halp}.

We now reobtain the conductance peaks from the analysis of 
the disk partition function, illustrating the method that
will be extensively used in the following sections.
The disk partition function for $\nu=1/p$ (\ref{z-disk})
including the area deformation reads, up to  irrelevant factors:
\be
\theta_\l=K_{\l}(\t,\z;p)=  
\frac{1}{\eta}\sum_{n=-\infty}^\infty \exp\left[ i2\pi 
\left(\t\frac{(np+\l-\s)^2}{2p}+\z \frac{np+\l}{p} \right)
\right] .
\label{k-fun}
\ee
Consider the Hall droplet without quasiparticles in the bulk 
corresponding to $\l=0$:
from the character $K_0$ (\ref{k-fun}), we can extract the energies 
and charges of the electron excitations
as the factors multiplying $\t$ and $\z$, respectively. 
Upon deformation of the dot area, the ground state energy, 
$E\sim \s^2/2p$, $Q =0$, and that of the one-electron state,  
$E\sim (\s-p)^2/2p$, $Q=1$, become degenerate at the midpoint $\s=p/2$,
leading again to (\ref{peak1}). 
In the presence of quasiparticles in the bulk with 
charge $Q=-\a/p$, one should repeat the analysis using the
partition functions $\theta_\a$ (\ref{z-disk}): one obtains the same 
peak separations, because the energies in the different $\l$ sectors 
are related by global shifts of $\s$. 

We see that the disk partition function is rather convenient for
studying the Coulomb blockade: its decomposition into sectors 
clearly indicates the allowed electron transitions and
considerably simplifies the analysis of more involved theories of
the following sections.
The strategy will be to first obtain the modular invariant
$Z_{\rm annulus}$ and then analyze the deformed energy spectra within each
disk sector.
 
We now discuss the peaks for variations of the magnetic field  $\D B$.
The chiral Luttinger RCFT is sensitive to flux changes, 
$\d= \D \f/\f_o =S\D B/\f_o$, 
as follows \cite{cz}: the partition functions are deformed by 
$K_\l\ \to\ K_{\l+\d}$, in (\ref{k-fun}) with $\s=0$ , i.e. the energies and
charges of all states are changed.  For one quantum of flux, $\d=1$,
each $\l$-sector ($\theta_\l$) goes into the following one: this is the
spectral flow discussed in the previous section.  For example, the
ground state $\theta_0$ becomes the one-anyon sector $\theta_1$, meaning
that the higher $B$ has induced a quasi-hole of charge $-1/q$ in the
bulk and the edge states have adjusted correspondingly \cite{geo2}.

As a matter of fact, the dependence of conductance peaks on variations of $B$
is the same as that for variations of $S$, because the edge energies
depend of the combined variable $(\d -\s)$,
\be
E_{\l,\s,\d}(n) = \frac{v}{R}
\frac{\left(\l+p n+\d -\s \right)^2}{2p} \ .
\label{double-var}
\ee
It is nevertheless important to stress the difference between  
$S$ and $B$ deformations: the former is a nonrelativistic electrostatic
effect, causing no change in the quantization of the Luttinger liquid
theory. The latter is the relativistic effect of the chiral anomaly,
leading to  charge nonconservation at the edge and the spectral flow
\cite{cdtz}.

\subsection{Partition functions of hierarchical states} 

The multi-component generalization of the Abelian theory of
the previous section is obtained as follows 
\cite{hiera}:
the electron fluid is assumed to have
$m$ independent edges, each one described by a Luttinger liquid,
altogether yielding the $\u1^{\otimes m}$ affine algebra \cite{cft}. 
Its  representations are labeled by a vector 
of (mathematical) charges $r_a, a=1,\dots,m$,
which spans an $m$-dimensional lattice, ${\vec r}=\sum_i{\vec v}^{(i)} n_i$, 
$n_i \in  \Z$, that is the closed set for the addition 
of charge vectors (the Abelian fusion rules).
The physical charge is a linear functional of ${\vec r}$ and the Virasoro
dimension is a quadratic form, both parameterized by the 
metric of the lattice,
$K^{-1}_{ij}= {\vec v}^{(i)}\cdot{\vec v}^{(j)}$.
The spectrum is therefore:
\ba
Q & = & \sum_{i,j=1}^m\ t_i\ K^{-1}_{ij}\ n_j\ , \qquad n_i\in \Z \ ,\nl
L_0 & = & {1\over 2}\ \sum_{i,j=1}^m\ n_i\ K^{-1}_{ij}\ n_j\ , \nl
\nu & = & \sum_{i,j=1}^m\ t_i \ K^{-1}_{ij}\ t_j\ .
\label{mabel}\ea
In these expressions, $K$ is an arbitrary symmetric matrix of couplings,
with integer elements, odd on the diagonal, due to
requirement that the spectrum contains $m$ electron-like excitations
\cite{hiera}. 
The vector $\vec t$ can be set to, $\vec{t}=(1,\dots,1)$, 
in a standard basis \cite{hiera}.
The spectrum (\ref{mabel}) is very general due to the many free
parameters in the $K$ matrix: these can actually be chosen to 
reproduce the results of all known hierarchical constructions of wave
functions \cite{wen}. 

The prominent Jain hierarchical states \cite{jain}, 
with filling fraction $\nu=m/(ms\pm 1)$, $m=2,3,\dots$, $s=2,4,\dots$,
where shown to correspond 
to the matrices $K_{ij}= \pm\delta_{ij} + s\ P_{ij}$, where 
$s>0$ is an even integer and $P_{ij}=1,\ \forall\ i,j=1,\dots,m$
\cite{hiera}.
The Jain spectrum is:
\ba
\nu& =& {m\over ms \pm 1}\ , \qquad s >0 \ {\rm even\ integer}\ ,
\qquad c = m\ , \nonumber\\
Q & = & {1\over ms \pm 1}\ \sum_{i=1}^m n_i \ ,\nonumber\\
L_0 &=& \pm {1\over 2}\left( \sum_{i=1}^m n^2_i -
{s\over ms \pm 1}\left( \sum_{i=1}^m n_i \right)^2 \right) \ .
\label{wspec}
\ea
(Let us first disregard the case with the minus signs, corresponding 
to antichiral neutral excitations).
The spectrum (\ref{wspec}) is rather peculiar because it
contains $m(m-1)$ {\it neutral} states with unit Virasoro dimension,
$(Q,L_0)=(0,1)$. By using a bosonic free field construction, 
one can show that these are chiral currents $J_{\vec\beta}$,
that can be labeled by the simple roots of $SU(m)$ and  
generate the $\u1\otimes\suem$ affine algebra at level one ($c=m$)
\cite{hiera}\cite{w-min}.

The annulus partition functions for multicomponent Luttinger liquids 
were obtained in \cite{cz}: 
we recall their expressions, first for general $K$-matrix theories
and then for the Jain hierarchy.

As in the previous section, the $U$ modular condition is the most
relevant one. In order to solve it, we first group the states with
integer-spaced charges in the $K$ lattice (\ref{mabel}).
These are clearly parameterized by
$ \vec{n}=K\vec{\ell}+ \vec{\lambda}\ $, with $\ \vec{\ell} \in {\Z}^m$.
Since $K$ is an integer matrix, there is a finite number of 
$\vec{\lambda}$ values (the sectors of the RCFT), belonging to 
the quotient of the $\vec{n}$ lattice by the $\vec{\ell}$ lattice:
\be 
\vec{\lambda} \in {{\Z}^m \over K \ {\Z}^m} \ .
\label{mlatt}\ee 
As in section 2.2,  the $\u1$ characters in each sector sum up into 
$m$-dimensional generalization of the theta functions (\ref{thetaf}): 
\ba
\!\!\!\!\theta_{\vec{\lambda}} &=&
{\rm e}^{\disp -{\pi\ t^T K^{-1}t}\ {\left(\I\zeta\right)^2\over\I\tau}}
\times\nl
& &\!\!
\frac{1}{\eta(q)^m}\ \sum_{\vec{\ell}\ \in\ {\Z}^m}
{\rm e}^{\disp i2\pi \left[
\frac{\tau}{2} \left(K\vec\ell+\vec\l\right)^T 
K^{-1}\left(K\vec\ell+\vec\l\right)
+\zeta\ \vec t^T \left(\vec\ell + K^{-1}\vec\l\right) \right] }. 
\label{mthetaf}\ea
Their $T^2,S,U,V$ transformations are straightforward generalizations
of (\ref{chitr}) and can be found in \cite{cz}: again, the characters
(\ref{mthetaf}) carry a finite-dimensional unitary projective
representation of the modular group. The dimension of representation
is $ |\det K| $ from (\ref{mlatt}) and matches the Wen topological order.

The $U$ invariance of the annulus partition function, written
as a sesquilinear form of the characters (\ref{mthetaf}), implies the equation
$\ t^T K^{-1}(\vec{\l}+\vec{\mu})\ \in {\Z}$
for the left and right weights. Its solutions depend on the specific 
form of $K$; here, we shall only discuss the diagonal 
solution, $\vec{\lambda}+\vec{\mu}=0$, that also solves
the other $(T^2,S,V)$ conditions.
Therefore, the modular invariant partition function is:
\be
Z^{\u1^{\otimes m}}_{\rm annulus}
\ = \ \sum_{\vec{\lambda}\in \Z^m/K\Z^m}\ \theta_{\vec{\lambda}}\ 
\ov\theta_{\vec{\lambda}} \ .
\label{zem}\ee

In the Jain hierarchical case, the $\suem$ symmetry can be used to
rewrite sums of $(m-1)$-dimensional $\u1$ characters into generalized
characters pertaining to representations of this extended symmetry: indeed,
all states in the $(m-1)$-dimensional sectors have integer-spaced dimensions.
The $\suem$ representations and characters were described in
Ref.\cite{itz}: there are $m$ highest weight representations,
corresponding to completely antisymmetric tensor representations of
the $SU(m)$ Lie algebra. 
They are characterized by an additive quantum number modulo $m$, the
so-called $m$-ality, $\alpha=1,\dots,m$: thus, the $\suem$ fusion rules are
isomorphic to the ${\Z}_m$ group.
Since the $\suem$ excitations are neutral, we shall
need their characters for $\zeta=0$, denoted by $\chi_\a(\t,0)$:
they clearly obey $\chi_\a(\t,0)=\chi_{\a\pm m}(\t,0)$.
The Virasoro dimension of $\suem$ representations is:
\be
h_\a={\alpha(m-\alpha)\over 2m}\ ,\qquad \alpha=0,\dots,m-1\ ,
\label{suem-dim}
\ee
The explicit form of the $\suem$ characters will not be
necessary in the following, 
but their leading $|q|\to 0$ behavior:
\be 
\c_\a(\t)\sim\left({m\atop \a} \right) \exp \left[
i2\pi\t \frac{v_n}{v}\left( \frac{\a(m-\a)}{2m} -
\frac{m-1}{24} \right)\right] + \cdots , \qquad
\a=0,1,\dots,m-1 \ .
\label{chi-sum} 
\ee 
In this expression we introduced a different Fermi velocity  $v_n$ for
neutral edge excitations, whose experimental value is expected
to be, $v_n\sim v/10$ \cite{halp}.

The modular transformations of $\suem$ characters are \cite{itz},
\ba
T^2  :\ \chi_\alpha\left(\tau+2\right) &=&
{\rm e}^{\disp i2\pi\left({\alpha(m-\alpha)\over m}-{m-1\over 12}\right)}
\ \chi_\alpha\left(\tau\right)\ ,\nl
S:\ \chi_\alpha \left(-{1\over\tau}\right) & = &
{1\over\sqrt{m}}\ \sum_{\alpha^\prime=1}^m\
{\rm e}^{\disp -i2\pi{\alpha\alpha^\prime\over m} } \
\chi_{\alpha^\prime}\left(\tau\right)\ ,
\label{chim}\ea
while $U,V$ do not act on neutral states.

Furthermore, the $\u1$ states in $\u1\otimes\suem$ theories 
can be described by the ${\Z}_p$ characters (\ref{thetaf}) 
of section 2.2, with free parameter $p$, 
$K_\lambda(\tau,\zeta;p)$, with $\ \lambda=1,\dots,p$.
In summary, the modular invariance problem can be reformulated
in the two-dimensional basis of $K_\l \c_\a$ characters. 
The derivation of the partition functions in this basis 
\cite{cz} is rather instructive for the analysis of 
Read-Rezayi states of the next section.

The Jain spectrum (\ref{wspec}) can be
rewritten in a basis that makes apparent the decomposition into 
$\u1$ and $\suem$ sectors: upon substitution of,
\be
n_1 = l+\sum_{i=2}^{m} k_i \pm \a\ ,\qquad
n_i = l - k_i\ ,\ i=2,\dots,m\ ,
\label{chba}\ee
where $l,k_i\in{\Z}$ and $\alpha $ is the $\suem$ weight, we find:
\ba
\nu={m\over ms \pm 1}\ ,\qquad 
\ Q & = & \frac{ ml \pm \alpha}{ms \pm 1}\ ,\nl
L_0 &=& {\left(ml \pm\alpha\right)^2 \over 2 m(ms \pm 1)} \pm 
\frac{\alpha(m-\alpha)}{2m} +\ r\ , \quad r \in{\Z}\ .
\label{usspec}\ea
Consider these formulas with the upper signs, the other choice
will be discussed later.
One recognizes the $\suem$ contributions to $L_0$ (\ref{suem-dim}),
while the $\u1$ part identifies the parameters of $K_\lambda$ as
$p=m\hat p$, $\hat p=(ms+ 1)$ and 
$\lambda=ml +\alpha\ $ mod $(m\hat p)$; note that $\hat p= 1$ mod $m$ and
$\lambda=\alpha$ mod $m$, and that $\hat p, m$ are coprime numbers 
$(\hat p,m)=1$.
The normalization of $\z$ is chosen for $K_\l$ (\ref{thetaf}) 
to reproduce the spectrum (\ref{usspec}): this is,
$K_\l\left(\tau,m\zeta;p\right)$.

The tensor characters $K_\l \c_\a$ form a $(m^2\hat{p})$ dimensional basis
on which the $S$ transformation act as the $\Z_{m\hat{p}}\times\Z_m$ Fourier
transform. The analysis of the spectrum shows that charged and neutral 
excitations are paired by the condition $\l=\a$ mod $m$: therefore,
we should choose a subspace of tensor characters that obeys this condition
and is closed under modular transformations.
The dimension of this subspace is the topological order $\hat p$.
A well-known trick to reduce the space of the discrete Fourier transform
by a a square factor $m^2$  \cite{ade} is to
consider the following $m$-term linear combinations of characters 
obeying $\l=\a$ mod $m$ \cite{cz},
\be
{\rm Ch}_\lambda\left(\tau,\zeta\right) = \sum_{\b=1}^m\  
K_{\lambda+\b\hat p}\left(\tau,m\zeta;m\hat{p}\right)\ 
\chi_{\lambda +\b\hat p\ {\rm mod}\ m}\left(\tau,0\right)\ , 
\qquad \lambda=1,\dots, m\hat p\ .
\label{ddiag}\ee
These characters satisfy ${\rm Ch}_{\lambda+\hat p}={\rm Ch}_\lambda$; 
thus, there are $\hat p$ independent ones, which can be chosen to be
(due to $\hat p=1$ mod $m$):
\be
\theta_a= {\rm Ch}_{m a}\ , \qquad a=1,\dots,\hat p \ ,\qquad
\hat p = ms+1 \ .
\label{theta-a}
\ee

One can check that these generalized characters
$\theta_a$ carry a $\hat p$ dimensional representation of the modular
transformations $(T^2,S,U,V)$, with
$S_{ab}\propto\exp\left(i2\pi\ mab/\hat p\right)/\sqrt{\hat p}\ $.
The modular invariant annulus partition function is therefore the
diagonal expression in these characters:
\be
Z^{\u1\times\suem}_{\rm annulus}=
\sum_{a=1}^{\hat p}\ \theta_{a}\ \ov{\theta}_{a}=
\sum_{a=1}^{\hat p}\  
\left(\sum_{\a=1}^m\ K_{ma+\a\hat p}\ \chi_\a \right)
\left(\sum_{\b=1}^m\ \ov K_{ma+\b\hat p}
       \ \ov\chi_\b \right)\ .
\label{dsuem}\ee
Indeed, the characters $\theta_{a}$ can be shown to be
equal to the $\chi_{\vec{\lambda}}$ in (\ref{mthetaf}) for 
$\ K=1+sC\ $, once the corresponding $\u1$ charges are identified 
\cite{cz}.

The $V$ transformation,
\be
V \ :\ \theta_a\left(\tau,\zeta+1\right) =
{\rm e}^{\disp -i2\pi {m\over \hat p}
\left(\R \zeta +\R {\tau\over 2}\right)}
\ \theta_{a+1}\left(\tau,\zeta \right)\ ,
\label{stillv}\ee
shows that the minimal transport of charge between the two edges is 
$m$ times the elementary fractional charge; this is the smallest spectral
flow among the states contained in (\ref{ddiag}) which keeps $\b$
constant, namely which conserves the $\suem$ quantum number carried by
the neutral excitations. The Hall current, $\nu=m/\hat p$, is thus 
recovered.
 
We now give the partition functions for Jain states with charged and
neutral excitations of opposite chiralities
on each edge, corresponding to $\nu=m/(ms -1)$ in (\ref{wspec}) \cite{cz}.
The chirality of the neutral excitations can be switched by a 
simple modification of the characters (\ref{ddiag}),
\be
{\rm Ch}^{(-)}_\lambda = \sum_{\b =1}^m\  
K_{\lambda+\a\hat p}\left(\tau,m\zeta;m\hat{p}\right)\ 
\ov\chi_{\lambda +\b\hat p\ {\rm mod}\ m}\ , 
\qquad \lambda=1,\dots, m\hat p=m(ms-1)\ .
\label{antic}\ee
This gives again a representation of the modular group for $\hat p=ms-1>0$. 
The $U$ condition implies the partition function: 
\be
Z^{(-)}_{\rm annulus}=\sum_{a=1}^{\hat p}\  
\left(\sum_{\a =1}^m\ K_{ma+\a\hat p}\ \ov\chi_\a \right)
\left(\sum_{\b =1}^m\ \ov K_{ma+\b\hat p}\ \chi_\b \right)\ .
\label{dcsuem}\ee
that is modular invariant and reproduces the spectrum 
(\ref{wspec}) for $\nu=m/(ms-1)$.

\subsection{Coulomb peaks in hierarchical states} 

The following disk partition functions can be extracted from 
the annulus expressions (\ref{dsuem}), 
\be
Z_{{\rm disk},a}^{\u1\otimes\suem} = \theta_a 
=\sum_{\b=1}^m K_{ma+\b \hat p} (\t,m\z;m\hat p)\ \c_\b(\t)\ 
\ , \quad a=1,\dots,\hat p,
\label{z-hier} 
\ee
where $\hat p=ms+1$.
The novelty w.r.t. the Laughlin case of section 3.1, is that 
each sector contains combined charged and neutral excitations, that are 
described by the $K_\l$ (\ref{k-fun}) and $\c_\a(\t)$ (\ref{chi-sum}) 
characters, respectively.
For example, in the $\nu=2/5$ case, $(m=2, \hat p=5)$,
there are two neutral characters 
that combine with ten charged ones to obtain the following 
five sectors: 
\ba 
\theta_0 &=& K_0 (\t,2\z;10)\ \c_0 +K_5(\t,2\z;10)\ \c_1\ ,\nl 
\theta_{\pm 1} &=& K_{\pm 2} (\t,2\z;10)\ \c_0 +
K_{5\pm 2} (\t,2\z;10)\ \c_1\ ,\nl 
\theta_{\pm 2} &=& K_{\pm 4} (\t,2\z;10)\ \c_0 +
K_{5\pm 4} (\t,2\z;10)\ \c_1\ . 
\label{z-ex} 
\ea 
We now search for degeneracy of energy levels
differing by the addition of one electron, $\D Q=1$. 
Consider for definiteness the $\nu=2/5$ case 
without any bulk quasiparticle, i.e. $\theta_0$ above. 
From the expressions (\ref{k-fun}, \ref{z-hier}, \ref{z-ex}), 
one finds that the first term 
$K_0$  resumes all even integer charged excitations, while 
$K_5$ the odd integer ones. Therefore, the first conductance peak is
found when the lowest energy state in $K_0\ \c_0$, i.e. 
the ground state, with $E=(v_c/R) (2\s)^2/20$, $Q=0$, 
becomes degenerate with the lowest one in $K_5\ \c_1$, with 
$E=(v_c/R)(-5 +2 \s)^2/20 +v_n/4R $, $Q=1$. 
The next peak occurs when the latter becomes degenerate with  
the first excited state ($Q=2$) in $K_0\ \c_0$, and so on. 
Owing to the contribution of the neutral energy in $\c_1$ 
(cf. (\ref{chi-sum})), the level matching is not midway and there is 
a bunching  of peaks in pairs, with separations
$\s =5/2 \mp v_n/2v$ (see Fig. \ref{lev-fig}(b)).
Note that in the previous energies we modified $\s\to 2\s$,
(cf. (\ref{k-fun})), in order to respect the flux-charge relation, 
$\D Q =\nu \D\f/\f_o$.

For general $m$ values, the result can be similarly obtained by comparing 
the energies in consecutive pairs of terms,
$\b$ and $\b+1$, in $\theta_0$ (\ref{z-hier}). 
One finds:
\be
\D \s_\b=\s_{\b+1}-\s_\b=\frac{\hat p}{m}+ \frac{v_n}{v}\left(
h_{\b+2}-2 h_{\b+1}+h_\b \right)\ .
\label{ab-diff}
\ee
where $h_\b$ are the $\suem$ dimensions (\ref{suem-dim}).  Since they are
quadratic in $\b$, the discrete second derivative in
(\ref{ab-diff}) is constant, except for one value at the border of
the $m$ period.  This implies that there are groups of 
$m$ equally spaced peaks, with a larger spacing between groups.

An important fact shown by the $\wh{SU(m)}_1$ character (\ref{chi-sum}), 
is that the low-lying neutral states occur with characteristic multiplicities
$d_\b=\left({m\atop \b}\right)$.
This means that $d_\b$ degenerate states are simultaneously
made available for the $\b$-th electron to tunnel into the droplet. 
These multiplicities can be easily understood in a classical model 
of $m$ superposed fluids, where the one-electron excitation is
$m$ times degenerate, the two-electron one is $m(m-1)/2$ times and so on.

Summarizing, in the Jain hierarchical states, $\nu=m/\hat p$, the peak pattern
is the following: the separation $\D \s_k$ between the $k$-th and $(k+1)$-th
peaks and the level multiplicity $d_k$ read ($\s=B\D S/\f_o$):
\ba
\D\s_k \! &=& \!  \frac{\hat p}{m}- \frac{v_n}{v}\frac{1}{m}\ ,
\qquad\qquad d_k=\left({m\atop k}\right), \  k=1,\dots,m-1,
\cr
\D\s_m \!  &=& \!  \frac{\hat p}{m}+ \frac{v_n}{v}\frac{m-1}{m}\ ,
\ \qquad d_m=1\ . 
\label{peak} 
\ea 
The pattern repeats with periodicity $m$.  It is independent of
the presence of quasiparticles in the bulk, because the corresponding
sectors, $\theta_a$, $a\neq 0$, have linearly shifted energies w.r.t
those of $\theta_0$ (cf.(\ref{z-hier})) \cite{cgz}.  
Note that the bulk quasiparticles have the same multiplicities: for example, at
 $\nu=2/5$ there are two quasiparticles with $Q=1/5$
(cf. (\ref{z-ex})).  In presence of quasiparticles with 
multiplicity $\left({m \atop k-1} \right)$, the sequence of peaks 
(\ref{peak}) starts from $\D\s_k$ (instead of $\D\s_1$) and 
goes on periodically.
We remark that the results (\ref{peak}) can also be obtained from 
the analysis of the $m$-dimensional lattice of excitations (\ref{wspec}), 
where the multiplicities $d_k$ are found by counting 
the shortest vectors with integer charge $k$.

\begin{figure}[t]
\begin{center}
\includegraphics[width=6cm]{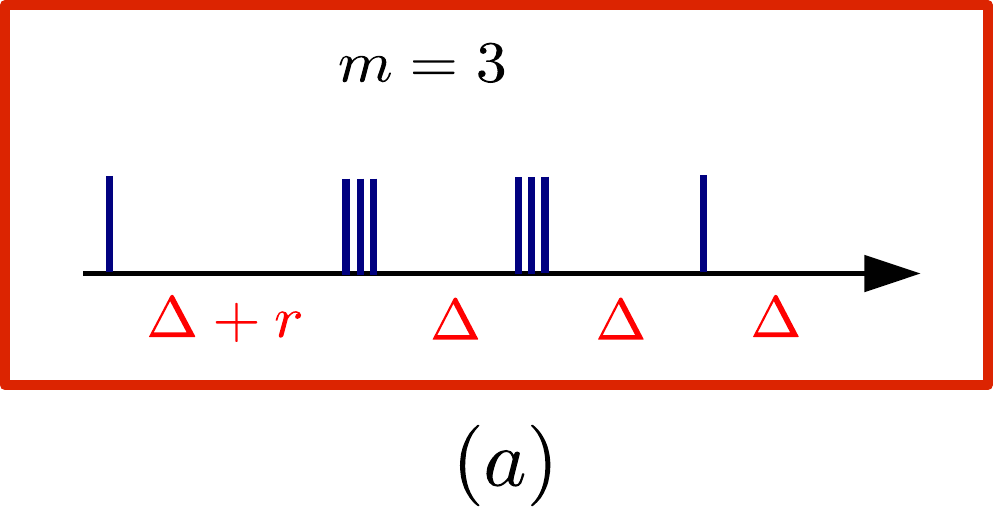}
\hspace{.5cm}
\includegraphics[width=7.5cm]{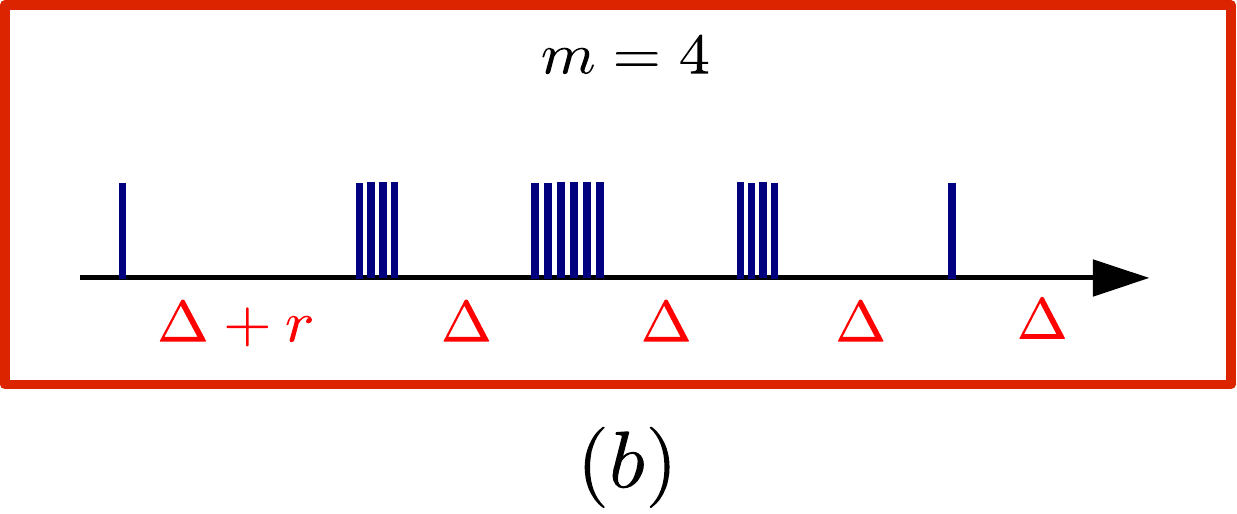}
\end{center}
\caption{Coulomb peaks in $m=3$ and $m=4$ Jain hierarchical states; 
the extra separation is $r=v_n/v\sim 0.1$.}
\label{jain-peaks}
\end{figure}

The multiplicities $d_\b$ of low-lying edge excitations could be
experimentally observed in the Coulomb peaks as follows. We expect that the
degeneracy is broken by finite-size effects, such as perturbations of the CFT
by irrelevant operators, that could be non-negligible for small droplets, with
sizes of the order of one micron, where Coulomb blockade effects are expected
to be found \cite{cb-exp}.
(Further level splittings could arise for unequal values of the
$(m-1)$ neutral modes velocities.)
If the symmetry breaking is small, the levels have a fine structure
and the electrons tunneling into them yield peaks at different,
slightly displaced distances: upon superposing several periods of
Coulomb peaks, one can observe the patterns shown in
Fig. \ref{jain-peaks} for $m=3,4$.  The single peak at the end of the
periods can be taken as a reference point for the superpositions and
as a measure of experimental errors.

In conclusion, the multiplicities of Jain hierarchical fluids are
observable if the experimental precision is sufficiently high and
the $\suem$ symmetry is not strongly broken.
Note that decays among the fine-structured levels
are possible and could take place on short time scales;
higher levels in the multiplets are nevertheless 
populated due to thermal fluctuations.
(Earlier discussions \cite{cgz} of experimental signatures 
of level multiplicities were not completely correct).

We now consider the Jain hierarchical states with mixed chiralities,
$\nu=m/(ms-1)$, described by the disk partition functions (\ref{antic}).
These were obtained by the replacement, $\c_\b\to\ov{\c}_\b$, that
does not affect the earlier discussion of energetics.  Therefore, the
formulae (\ref{peak}) also hold in these cases, upon replacing $\hat
p=ms-1$, and predict the same peak multiplicities.  

The dynamics of these $m$-composite edge theories was much discussed in
the literature, starting from the experimental result
\cite{chang}. Actually, the presence for $\nu=m/(ms-1)$ of neutral and charged
excitations of opposite chiralities on the same edge may allow for
destabilizing interactions, leading to edge reconstruction effects
\cite{agam}.
Several deformations of, or additions to the Luttinger
liquid Hamiltonian, have been put forward; as these break
the $\wh{SU(m)}_1$ symmetry and possibly the conformal symmetry, they
may lift the peak degeneracy.  
Therefore, it is interesting to find their predictions for the
Coulomb peak patterns and compare with
the present results of the $m$-component Abelian theory.


\subsection{Coulomb peaks in alternative hierarchical theories}

In this subsection we discuss the Coulomb peaks in 
two alternative theories of the Jain states:
the $\winf$ minimal models \cite{w-min},
introduced by Trugenberger and some of the authors, and the
three-fluid theory \cite{lf} by Fradkin and Lopez.  Both theories:

i) possess a reduced number of currents and more generally less
degrees of freedom w.r.t. the $m$ component Luttinger theory;

ii) are not rational CFT, i.e. their partition function are not
modular invariant. Nevertheless, these theories
are projections of RCFTs and the partition function can be found
by working out the reduction of degrees of freedom; the
Coulomb blockade peaks are then found by the same methods as before.

We first discuss the minimal $\winf$ models \cite{w-min}: these
theories were introduced by exploiting the main geometrical feature of
incompressible Hall fluids, which is the symmetry under
area-preserving diffeomorphisms of the plane, the local coordinate
transformations that keep the density
constant \cite{w-simm}.  The edge excitations can be seen as infinitesimal
area-preserving deformations of the Hall droplet , and can be
naturally described by the representations of the
symmetry algebra.  In the mathematical literature, this is called the
$\winf$ algebra and its representations on the circle, i.e. on the
edge CFT, have been completely classified \cite{kac}.

The mathematical results are the following: for generic parameters,
the $\winf$ representations are isomorphic to those of the
multicomponent Luttinger theory and their $m$ dimensional $K$
lattices.  However, for special lattices, there exist degenerate
representations, with a reduced set of excitations: as shown in
\cite{w-min}, these correspond one-to-one with the $\wh{SU(m)}_1$ symmetric
lattices of the Jain states discussed before.  For these theories, it
is thus possible to project out states of the Luttinger theory and
obtain the minimal $\winf$ models made of irreducible representations
only.  The main features of these models are \cite{w-min}:

i) the multiplicities due to the $SU(m)$ symmetry are completely
eliminated; in particular, there is only one conserved $U(1)$ current,
the electric current, a single electron state (not $m$) and no 
further degeneracies in the spectrum. The
charges and conformal dimensions are still given by the formulas 
(\ref{wspec}), but the integer labels are constrained within the wedge,
\be
n_1\ge n_2\ge \dots \ge n_m\ ;
\label{wedge}
\ee 
ii) the conformal symmetry is maintained, but the partition
function is not modular invariant.  The projection from the $m$
component Luttinger liquid to the minimal models can be realized
by introducing a non-local, $\suem$ breaking interaction in 
the Luttinger Hamiltonian, that commutes with Virasoro \cite{h-red}.
This term is diagonal in the $SU(m)$ basis and gives higher energies to
the unwanted states; at infinite coupling the projection is realized
leading to the $\winf$ minimal theory.  
The non-locality of the interaction term in the Hamiltonian
explains the lack of modular invariance in these theories.

The conductance peaks in the minimal $\winf$ theories are easily
obtained: since the projection preserves the structure of the Hilbert
space of the $(c=m)$ Luttinger theory, it does not modify the structure of
disk partition functions (\ref{z-hier}), but only replaces the neutral
characters $\c_\a$ by other expressions whose leading terms
(\ref{chi-sum}) have no multiplicity factors \cite{huerta}:
\be 
\c^{\winf}_\a(\t)\sim \exp \left[
i2\pi\t \frac{v_n}{v}\left( \frac{\a(m-\a)}{2m} -
\frac{m-1}{24} \right)\right] + \cdots , \qquad
\a=0,1,\dots,m-1 \ .
\label{chi-min} 
\ee 
The absence of multiplicities can also be understood from the charge
lattice (\ref{wspec}) as due to the constraint (\ref{wedge}).  We
conclude that the $\winf$ minimal models predict conductance peaks in
Jain states with the same pattern (\ref{peak}) as the Luttinger
theory but without any multiplicity.

The Lopez-Fradkin theory of Jain's states is a variant of the multicomponent
Luttinger liquid and can be formulated in the
charge lattice approach introduced before: all Jain states,
$\nu=m/(ms+1)$, $s$ even, are described by the same
three-dimensional $K$ matrix and charge vector $t$,
\be
K= \left(
\begin{array}{ccc}
-s & 1 & o \\ 1  & m & 0\\ 0 & 0& 1
\end{array}
\right)\ , \qquad t=(1,0,0)\ .
\label{l-f}
\ee

The first component clearly describes the charged excitations; the other
two sectors are called topological, because their neutral excitations
do not propagate, i.e.  $v_n=0$. This choice modifies the time
scaling of the electron correlator: for $1/3 <\nu< 1/2$, the
multicomponent Luttinger theory predicts a constant exponent $\a=3$,
while the Lopez-Fradkin theory a varying one, $\a=1/\nu$, which is in
better agreement with the experimental results
\cite{chang}.  Note, however, that the $m$ component Luttinger theory
also predicts $\a=1/\nu$ in the limit $v_n=0$, while other experiments
would favor $0<v_n \ll v$ \cite{nico}.

Another feature of Lopez-Fradkin theory is that 
the excitations are described by a subspace of the three dimensional
charge lattice (\ref{mabel}): the integers
$(n_1,n_2,n_3)$ labeling excitations are constrained by the condition 
of physical states, $n_2=-n_3$. Therefore, the partition function
has the general charge-lattice form (\ref{zem}), but it is not 
modular invariant due to this constraint.

The Coulomb peaks in this theory are equidistant because the modulations are
proportional to $v_n/v=0$: moreover, there is a single electron excitation and
no level multiplicities because the two dimensional sublattice of
(\ref{l-f}) has no symmetries.  For the same reason, the Coulomb peaks would
be equally spaced even for $v_n>0$.

In conclusion, we have shown that three proposed theories for the Jain
states predict rather different patterns of conductance peaks that
would be interesting to test experimentally.  Other edge theories have
been proposed whose peak patters remain to be investigated \cite{agam}.


\section{Coulomb Blockade in Read-Rezayi states}

The Pfaffian ($k=2$) \cite{pfaff} and Read-Rezayi ($k=3,4,\dots$) \cite{rr}
theories are prominent candidates for describing spin-polarized Hall plateaux 
observed in the second Landau level \cite{kang}. 
The wave functions of these theories describe
electrons that first bound themselves into $k$-clusters and then 
form incompressible fluids. In particular, the $k=2$ state corresponds to a
two-dimensional p-wave superconductor \cite{green} and should be realized
at $\nu=5/2$.
These states are the present best candidates for observing excitations 
with non-Abelian statistics \cite{na-interf} and for manipulating them 
to realize the unitary transformations of quantum
computation, following the proposal of topological quantum computation 
\cite{tqc}.
Therefore, Coulomb blockade and interferometry of edge excitations
 are actively investigated in these states both theoretically 
and experimentally \cite{halp}\cite{stern}\cite{schou}\cite{cb-exp}.

The filling fractions are:
\be
\nu =2 + \frac{k}{kM+2}\ , \qquad k=2,3,\dots, \quad M=1,3,\dots
\label{nu-RR}
\ee
The RCFT description is based on a charged Luttinger field and the
neutral $\Z_k$ para\-fermion theory, $\u1\times {\rm PF}_k$, 
 with central charge $c=1 + \frac{2(k-1)}{k+2}$ \cite{z-k}.
The best observable plateaus correspond to $M=1$ but we shall provide
formulas for the general case.
From the wave function constructions and earlier analyses, the following 
spectrum has been found \cite{cgt2}:
\ba
Q &=& \frac{m}{kM+2} +s \ ,\quad s \in \Z\ , 
\nl
L_0 &=& \frac{\left( m+s(kM+2)\right)^2}{2k(kM+2)} + h^\ell_m \ ,
\label{pf-spec}
\ea 
where $h^\ell_m$ is the dimension of parafermion states to be
described momentarily.
The topological order, i.e. the number of sectors of the RCFT,
is: 
\be
\frac{(k+1)(kM+2)}{2}\ ,
\label{pf-top}
\ee
and is obtained as follows.  The fractional charge, $Q=\l/(kM+2)$,
implies $(kM+2)$ Abelian sectors, and the $\Z_k$ parafermions
possess $k(k+1)/2$ sectors: the $\Z_k$ parity rule \cite{cgt2},
\be
\l = m \quad {\rm mod\ } k\ ,\qquad \l \ {\rm mod\ } (kM+2) \ , 
\qquad m \ {\rm mod\ } 2k\ ,
\label{pf-rule}
\ee
relates the fractional charge $\l$ to the parafermion charge $m$, 
leading to the multiplicity of $(\l,m)$ pairs in (\ref{pf-top}).
One physical motivation for the parity rule is the requirement
of locality (analyticity) of all excitations w.r.t.  electrons
in the wave functions (cf. (\ref{estat})) \cite{cgt2}. 

\subsection{Partition functions}

The partition functions of Read-Rezayi states have been already obtained
in \cite{cgt2} from a 
physically motivated construction involving the projection from
the Abelian theory with $\u1\times\wh{SU(k)}_1\times\wh{SU(k)}_1$
symmetry. In particular, the $\Z_k$ parafermionic theory for neutral
excitations was described by the coset construction
${\rm PF}_k=\wh{SU(k)}_1\times\wh{SU(k)}_1/\wh{SU(k)}_2 $.

In the following, the annulus partition functions will be found by
solving the modular conditions as in section 3.2; the parafermions
will be described by the standard coset, 
${\rm  PF}_k=\wh{SU(2)}_k/\u1_{2k}$, that characterizes their sector by the
quantum numbers $(\ell, m)$, equal to twice the spin and spin
component, respectively \cite{qiu}.

\begin{figure}[t]
\begin{center}
\includegraphics[width=5cm]{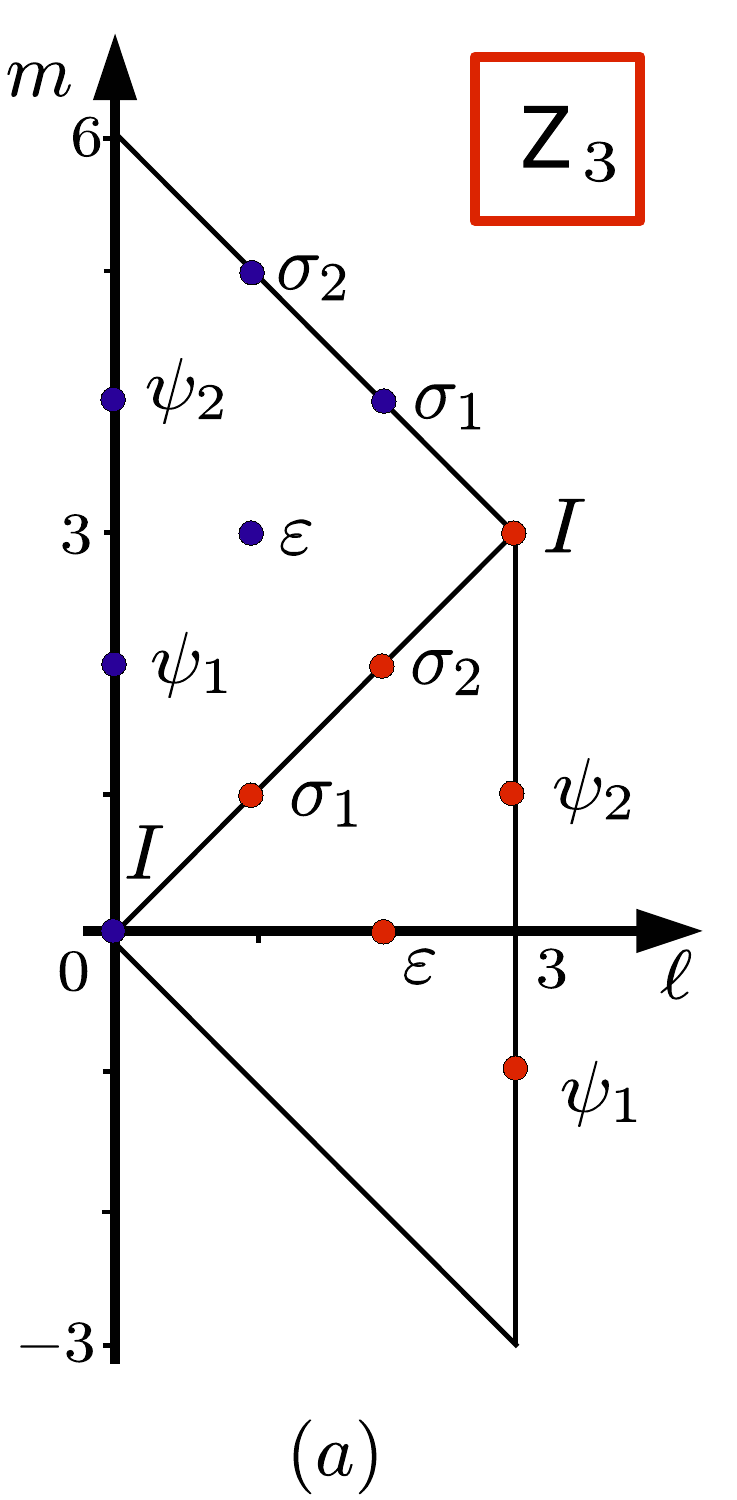}
\hspace{4cm}
\includegraphics[width=5cm]{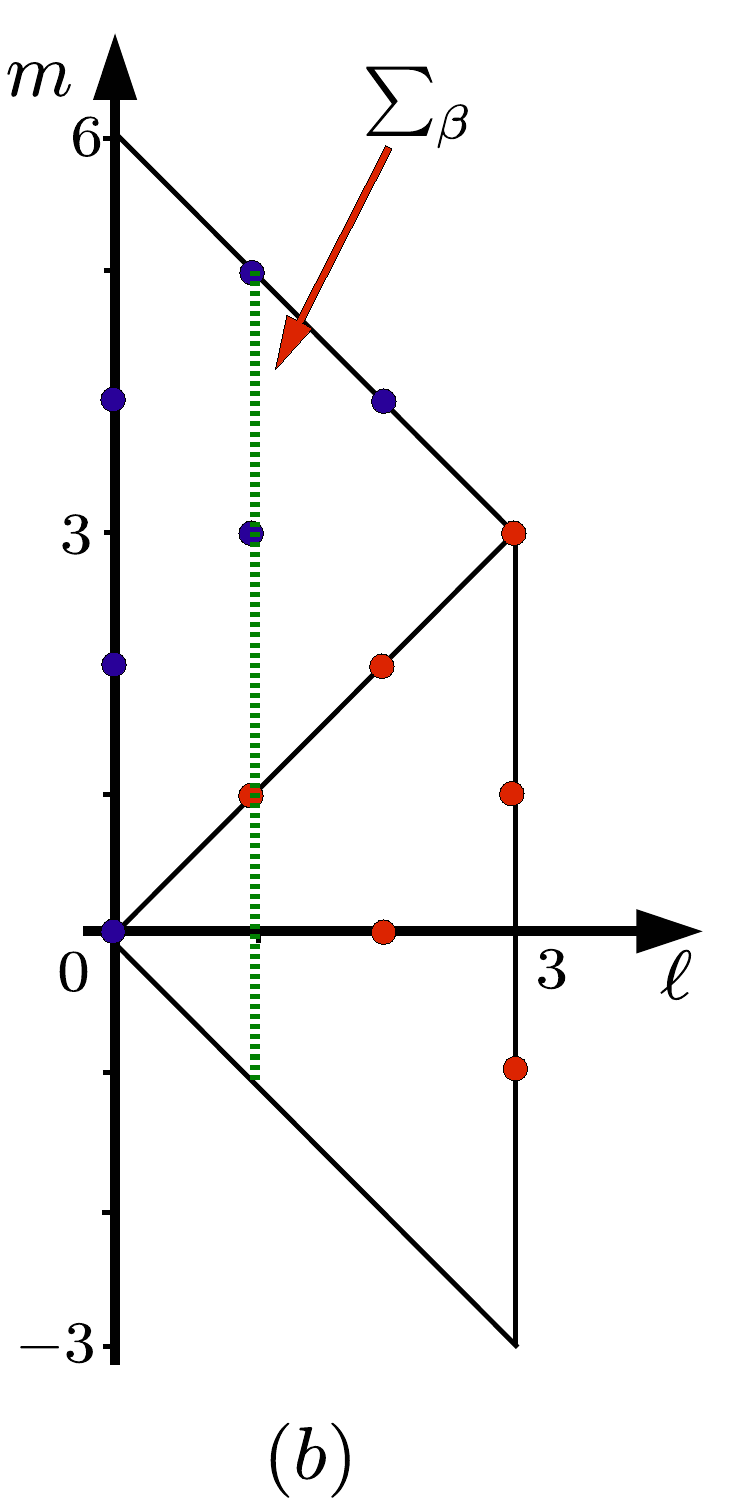}
\end{center}
\caption{Diagram of $\Z_3$ parafermion fields: (a) field symbols; 
(b) on the green line, fields involved in $\ell=1$ sectors of
the $k=3$ Read-Rezayi partition function.}
\label{pf-fig}
\end{figure}

The dimensions of parafermionic fields $\f^\ell_m$ are given by:
\be
h_m^\ell =\frac{\ell(\ell+2)}{4(k+2)}-\frac{m^2}{4k}\ ,
\qquad \ell=0,1,\dots,k, \qquad -\ell <m\le \ell, \quad \ell=m
{\rm \ mod\ }2\ .
\label{hlm}
\ee
For example, the $\Z_3$ parafermion fields are shown in Fig. \ref{pf-fig}(a):
the coset construction implies that the $m$ charge is defined modulo
$2k$ \cite{qiu}, thus the fields are repeated once outside the fundamental 
$(\ell,m)$ domain (\ref{refl}) by the reflection-translation,
$(\ell,m)\to (k-\ell,m+ k)$,
\be
\f^\ell_m =\f^{k-\ell}_{m-k}\ , \qquad \ell=0,1,\dots,k,\qquad
l<m\le 2k-l\ .
\label{refl}
\ee

The fusion rules are given by the addition of the
$\wh{SU(2)}_k$ spin and $\u1_{2k}$ charge:
\be
\f_m^\ell\cdot\f_{m'}^{\ell'} =
\sum_{\ell''=|\ell-\ell'|}^{{\rm min}(\ell+\ell',2k-\ell-\ell')}\ 
\f^{\ell''}_{m+m'\ {\rm mod}\ 2k}\ .
\label{fus-RR}
\ee
In particular, the parafermions $\psi_j =\f_{2j}^0$, $j=1,\dots,k-1$, obey
Abelian fusion rules w.r.t. the index $j$ mod $k$.  In the Read-Rezayi states,
$\psi_1$ represents the neutral component of the electron and the fusion,
$\left(\psi_1\right)^k = I$, describes the $k$-clustering property
of the ground state wave function \cite{rr}.

In order to find the annulus partition function, we need the
expressions and modular transformations of parafermionic characters,
denoted by $\c^\ell_m(\t;k)$: they obey,
\be
\c^\ell_m=\c^\ell_{m+2k}=\c^{k-\ell}_{m+k} \ , \quad
m=\ell\ {\rm mod}\ 2,\qquad\ \ \c^\ell_m=0\, \quad m=\ell+1\ {\rm mod}\ 2\ .
\label{pf-char}
\ee
The transformations can be obtained from those of $\wh{SU(2)}_k$ and
$\u1_{2k}$ characters, respectively denoted by $\Th_\ell(\t;k)$ and 
$K_m(\t,0;2k)$, by using the coset relation \cite{qiu}:
\be
\Th_\ell(\t;k)=\sum_{m=-k}^{k-1} K_m(\t,0,2k)\ \c^\ell_m(\t;k)\ .
\label{pf-cos}
\ee
The Luttinger sectors $K_\l$ were described in section 3 and the
$\wh{SU(2)}_k$ characters and transformations are well known
\cite{cft}: the $\wh{SU(2)}_k$ $S$ matrix has a simpler form  
if we extend the domain of the
index $\ell$  to $L=\ell+1$ mod $N=2(k+2)$, by defining,
$\Th_L=\Th_{L+N}=-\Th_{-L}$ (note that $\Th_L=0$ for $L=0,N/2$) \cite{ade}.
In the extended domain the transformation reads:
$\Th_L(-1/\t)=-i\sum_{L'\ {\rm mod} \ N}
\exp\left(i2\pi LL'/N \right) \Th_{L'}(\t)/\sqrt{N}$.
It follows that:
\be
\c^L_m\left(-1/\t;k\right)=\frac{-i}{\sqrt{4k(k+2)}}
\sum_{L'\atop {\rm mod}\ 2(k+2)}\sum_{m' \atop {\rm mod}\ 2k}\
e^{-i2\pi\frac{ mm'}{2k}}\ e^{i2\pi\frac{ LL'}{2(k+2)}} 
\ \c^{L'}_{m'}(\t;k)\ .
\label{pf-s}
\ee

The $\Z_k$ parity rule (\ref{pf-rule}) should now be used to couple the
neutral characters $\c^\ell_m$  to the charged
ones, given by $K_\l(\t,k\z;k\hat p)$,
$\hat p=kM+2$, in agreement with the spectrum (\ref{pf-spec}).
As in the case of hierarchical states of section 3.2,
we should find the $\hat p (k+1)/2$ sectors of the Read-Rezayi theory
within the $K_\l \c^\ell_m$ tensor basis of dimension $k^2$ times larger.
It is thus natural to introduce analogous sums of $k$ terms,
\be
\theta_a^\ell =\sum_{\b=1}^k K_{a+\b\hat p}(\t,k\z;k\hat p)\ 
\c^\ell_{a+2\b}(\t;k) \ , \qquad 
\begin{array}{l}
a=0,1,\dots, \hat p-1,\ \hat p=kM+2,\\
\ell=0,1,\dots,k,\\
a = \ell\ {\rm mod}\ 2,
\end{array}
\label{pf-th}
\ee
that obey the parity rule (\ref{pf-rule}) and reproduce the spectrum 
(\ref{pf-spec}).
Their periodicity, $\theta^\ell_{a+\hat p}=\theta^{k-\ell}_a$, 
justifies the indicated
ranges for the indices and checks the value of the topological order.

The $S$ transformation of the generalized characters  $ \theta_a^\ell$
can be found by using the
previous formulas; after returning to the $\ell$ index, it reads:
\be
\theta^\ell_a(-1/\t) =
\sqrt{\frac{2}{(k+2)\hat p}}
\sum_{a'=1}^{\hat p}\sum_{\ell'=0}^k\ 
e^{-i2\pi \frac{aa' M}{2\hat p}} 
\sin\left(\frac{\pi (\ell+1)(\ell'+1)}{k+2}  \right)\ 
\theta^{\ell'}_{a'}(\t) ,
\ \ a=\ell \ {\rm mod}\ 2,
\label{pf-th-s}
\ee
and $ \theta^\ell_a(-1/\t) $ vanishes for $a=\ell+1$ mod $2$
(hereafter, we disregard the global phase $\propto\R(\z^2/\t)$ in the
characters).
The annulus partition function of Read-Rezayi states is given by 
the diagonal sesquilinear form,
\be
Z_{\rm annulus}^{\rm RR} =\sum_{\ell=0}^k\ \ 
\sum_{a=0 \atop a=\ell \ {\rm mod}\ 2}^{\hat p-1}
\ \left\vert\ \theta^\ell_a\ \right\vert^2\ ,
\label{pf-zann}
\ee
that solves the $(S,T^2,U,V)$ conditions of section 2.2. One can check that 
these partition functions are equal to those found in \cite{cgt2}. 

Let us give some examples. In the Pfaffian state, the $\Z_2$ parafermions
are the three fields of the Ising model:
$\f^0_0=\f^2_2=I$, $\f^1_1=\f^1_3=\s$ and $\f^0_2=\f^2_0=\psi$,
of dimensions $h=0,1/16,1/2$, respectively.
For $\nu=5/2$, i.e. $M=1$ in (\ref{nu-RR}), the Pfaffian theory possesses
$6$ sectors. The partition function is as follows: denoting
the neutral characters with the same symbol of the field
and the charged ones by $K_\l=K_\l(\t,2\z;8)$, $K_\l=K_{\l+8}$, 
with charge $Q=\l/4+2\Z$, we rewrite (\ref{pf-zann}) as,
\ba
Z_{\rm annulus}^{\rm Pfaffian} &=&
\left\vert  K_0 I+ K_4\psi\right\vert^2\ +
\left\vert  K_0\psi+K_4 I \right\vert^2\ +
\left\vert \left( K_1+ K_{-3}\right)\s\right\vert^2
\nl
&+&
\left\vert K_2 I +K_{-2}\psi \right\vert^2\ +
\left\vert  K_2\psi+K_{-2} I \right\vert^2\ +
\left\vert \left( K_3+ K_{-1}\right)\s\right\vert^2\ .
\label{ising-zann}
\ea
The first term describes the ground state and its electron
excitations, such as those in $K_4\psi$ with $Q=1+2\Z$; in the third
and sixth terms, the characters $K_{\pm 1}\s$ contain the basic
quasiparticles with charge, $Q=\pm 1/4$, and non-Abelian fusion rules
$\s\cdot\s=I+\psi$.  The other sectors are less familiar: the second
term contains a $Q=0$ Ising-fermion excitation (in $K_0\psi$) and the
4th and 5th sectors describe $Q=\pm1/2$ Abelian quasiparticles.

The $k=3$ Read-Rezayi state is also interesting because it is the simplest
of these systems that could perform universal quantum computations by
braiding non-Abelian quasiparticles \cite{tqc}.  It could be observable
at $\nu=13/5$, i.e $M=1$, and possibly at $\nu=12/5$ by charge conjugation.
The $6$ $\Z_3$ parafermion fields are listed in Fig. \ref{pf-fig}(a) with 
the corresponding $(\ell,m)$ labels
(the complete list of their quantum numbers can be
found in \cite{cgt2}).  The characters obey:
$\c^\ell_m=\c^\ell_{m+6}=\c^{3-\ell}_{m+3}$.  The charged sectors are
given by $K_a=K_a(\t,3\z;15)$, with charge $Q=a/5+3\Z$, and the
topological order is equal to $10$. The annulus partition function can
be written: 
\ba 
Z_{\rm annulus}^{\Z_3} &=&\sum_{a=0,\pm 2} \sum_{\ell=0,2}
\left\vert
  K_a \c_a^\ell +K_{a+5} \c_{a+2}^\ell +K_{a-5} \c_{a-2}^\ell
\right\vert^2 \nl
&&+\! \sum_{a=\pm 1} \sum_{\ell=1,3}
\left\vert 
K_a \c_a^\ell +K_{a+5} \c_{a+2}^\ell +K_{a-5} \c_{a-2}^\ell
\right\vert^2 \ .
\label{z3-ann}
\ea
As in the Pfaffian case, the basic quasiparticles are represented by spin
fields, $\s_1K_1$, $\s_2K_{-1}$, with smallest charges $Q=\pm 1/5$ and
Virasoro dimensions;
since these excitations have the smallest $SU(2)$
spin, $(\ell,m)=(1,\pm 1)$, we can interpret the index $\ell$ as counting the
number of quasiholes in the system.


\subsection{Coulomb blockade}

The analysis of degenerate energies and conductance peaks under
area variations $\D S$ can be obtained from the disk partition functions
(\ref{pf-th}),
\ba
Z_{\rm disk}^{(a,\ell)}&=&\theta_a^\ell =
\sum_{\b=1}^k K_{a+\b\hat p}(\t,k\z;k\hat p)\ \c^\ell_{a+2\b}(\t;k) \ , 
\nl
&& a=1,\dots, \hat p=(kM+2),\quad \ell=0,1,\dots,k,\quad
a = \ell\ {\rm mod}\ 2.
\label{pf-disk}
\ea
As seen in the previous examples, each $(\ell,a)$ sector involves the
parafermion fields of same $\ell$ value (along the vertical line in
Fig. \ref{pf-fig}(b)), each one associated to charges mod $k\Z$. Therefore,
adding one electron corresponds to going from $\b\to\b+1$ in the corresponding
sector $\theta^\ell_a$ (\ref{pf-disk}).
The distance between the conductance peaks is given by:
\be
\D \s_\b^\ell=\s^\ell_{\b+1}-\s^\ell_\b=\frac{\hat p}{k}+ 
\frac{v_n}{v}\left(
h^\ell_{a+2\b+4}-2 h^\ell_{a+2\b+2}+h^\ell_{a+2\b} \right)\ ,
\label{pf-diff}
\ee
that generalizes the earlier Abelian formula (\ref{ab-diff}).

The peak distances are modulated by the energies of neutral parafermions
$h^\ell_m$ through their discrete second
derivative. This is constant, $\D^2 h^\ell_m=-2/k$, up
to discontinuities at the boundaries of the domains in the $(\ell,m)$ plane,
which are the diagonals, $\ell=\pm m$, $m$ mod $2k$ (see Fig. \ref{pf-fig}(a).).
Whenever $a+2\b+2$ in (\ref{pf-diff}) stays on one diagonal, the 
result is $\D^2 h^\ell_m=1-2/k$; at the crossing of two diagonals, 
$(\ell,m)=(0,0), (k,k)$, mod $(0,2k)$, it reads $\D^2 h^\ell_m=2-2/k$.
Therefore, the peak patterns are the following,
\ba
&&\begin{array}{llc}
\ell =0,k &:\ \ \D\s=\left(\D+2r,\D,\cdots,\D\right), &
\ \ (k) \ \ \ \ \ {\rm groups},
\\
\ell =1,\dots,k-1 &:\ \ \D\s=\left(\D+r,\D,\cdots,\D+r,\D,\dots,
\D\right), & (\ell) (k-\ell)\ \ {\rm groups},
\end{array}
\nl
&&\qquad\qquad\qquad\qquad\qquad\qquad
\D = \frac{1}{\nu}-\frac{v_n}{v}\frac{2}{k}\ ,
\qquad  r=\frac{v_n}{v}\ .
\label{pf-patt}
\ea
In these non-Abelian Hall states, the peak patterns depend on the number 
of basic $\s_1$ quasiparticles in the bulk: for $\ell$ of them, there are
groups of $\ell$ and $(k-\ell)$ equidistant peaks separated by a larger gap,
$\D+r$, $r=v_n/v$.
The patterns are symmetric by $\ell\leftrightarrow (k-\ell)$ and
depends on the other quantum number $a$ only in the (irrelevant) starting
point of the sequence.
In particular, for the Pfaffian state $(k=2)$, the peaks group in pairs
when the number of bulk quasiparticles is even, and are equidistant
when it is odd, the so called ``even-odd effect'' \cite{halp}. 
The analysis of $\c_a^\ell$ characters also shows that there are no
degenerate states in parafermionic excitations, and thus the Coulomb
peaks have no multiplicities.

The peak patterns repeat periodically with period 
$\D\s=k/\nu=kM+2$, apart from the case $k$ even and $\ell=k/2$, where it is
halved. The same results were found by the direct analysis of
parafermion Hilbert space in \cite{stern}\cite{schou}.

The peak pattern in the ground state sector of Read-Rezayi states is
actually the same as in the Jain hierarchical Hall states of section
3.2, up to a factor of 2 in the neutral energies. However, the Abelian
case is characterized by specific peak multiplicities and is independent
of the sector, i.e. of bulk excitations.

\subsection{Peak patterns in the $(S,B)$ plane}

\begin{figure}[t]
\begin{center}
\includegraphics[width=10cm]{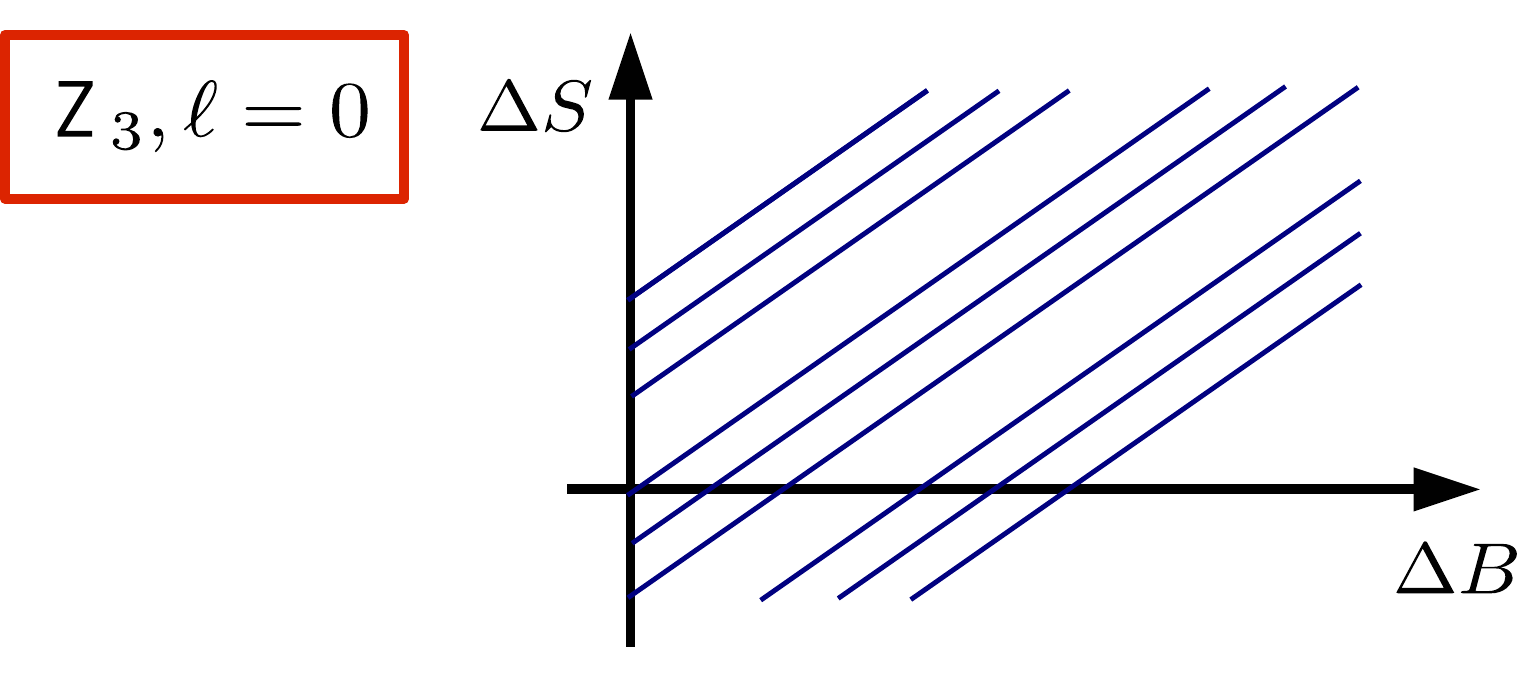}
\end{center}
\caption{Coulomb peaks in the area $(S)$ and magnetic field $(B)$ plane.}
\label{plane}
\end{figure}

We now analyze the peak patterns in the $(S,B)$ plane, i.e.
by simultaneous changes of area and magnetic field.
Let us first discuss $\D B$ changes at $\D S=0$: from section 3.1,
we recall that varying the field causes a chiral anomaly and
a drift of all charges in the theory. 
The addition of one flux quantum, $\d =S\D B/\f_o=1$
modifies the charged characters in the disk partition functions (\ref{pf-disk})
by $K_\l(\t,k\z;q)\to K_{\l+k}$, i.e. by a charge equal to the filling
fraction.
The spectral flow of Read-Rezayi sectors is therefore:
\be
\theta^\ell_a \ \to\ \theta_{a+k}^{k-\ell}=\theta^\ell_{a-2-(M-1)k}\ .
\label{pf-flow}
\ee
Namely, the sector $\ell$ goes into itself with a shift of the $a$ index:
the peak pattern already found for $S$ variations at $\D B=0$ 
repeats itself at $\D B=\f_o/S$ with an upward translation 
(see Fig. \ref{plane}). This continues for $\d=2,\dots$: eventually
the sector $\ell$ goes to the $(k-\ell)$ one, with the same
pattern, and then back to itself. 

This behavior of peaks in $B$ was also found for the Laughlin states
and actually holds in full generality: the peaks only depend on the 
combined variation $(\s-\d)$.
The proof is very simple: the magnetic field does not 
couple to the neutral characters, that are unchanged; since
they determine the peak modulations, the pattern remains the same.
The only effect is a rigid shift of the $\D S$ peak pattern at $\D B=0$.

In general, a magnetic flux induces a localized charge excitation
inside the Hall droplet, but this has to combine with a neutral 
excitation to form a physical state: in non-Abelian states, 
there might be several possibilities.
The result of the present analysis is that neutral parts are not
created and the $\ell$ sector does not change: there is a drift of states
within the same sector or the conjugate $(k-\ell)$.
One could naively think that $\D B>0$ would create the lowest charge
quasiparticle $\s_1$, but this would require a transition $(\D \ell,\D
m)=(1,1)$ that cannot be induced in isolated droplets within the
relativistic CFT description. Maybe it could be realized by
engineering a antidot inside the disk and by charging it, if further
non-relativistic effects are taken into account \cite{halp}.

In conclusion, the peak patterns in the $(S,B)$ plane are not
different from those on the $S$ axis. 
In ref. \cite{cb-exp}, it was observed that, for $\nu>2$, increasing $B$
causes a depletion of electrons in the second Landau level, 
that go into available states in the
first level; this is another effect that rigidly translate
upwards the peak patterns in the $B$ plane, i.e. it is of the same
sign as the spectral flow described here.


\subsection{Bulk-edge relaxation}

In reference \cite{schou}, a mechanism for relaxation of edge excitations
has been proposed. While the electric
charge is locally conserved at the edge, the neutral charge 
is (expected to be) only globally conserved: 
therefore, a (slow-time) process can be conceived in which neutral
excitations at edge and bulk fuse together thus achieving a lower energy
state at the edge.  In this mechanism, the electron added at the
boundary decays into another excitation with same charge but different
neutral content.  This process is possible whenever the theory possess
two or more excitations with same electric charge but different neutral
parts, among which the relaxation transition can take place.

In non-Abelian theories, there are necessarily many-to-one combinations
of neutral and charged parts.
Consider the following example of fusion rules:
\be
\f_1 \cdot \f_2 = \psi +\psi'\ .
\label{NA-fus}
\ee 
Both $\f_1,\f_2$ fields contain charged and neutral components,
and electric charge is conserved: thus, the fields
$\psi,\psi'$ have same electric charge but different neutral parts.
A possible relaxation at the edge is, $\psi \to \psi'$, by absorption
of one neutral bulk excitation, call it $\eps$,  
via the fusion, $\psi\cdot \eps=\psi'+\cdots$.
It is rather natural to expect that the $\eps$ field exists in the theory.

\begin{figure}[t]
\begin{center}
\includegraphics[width=6cm]{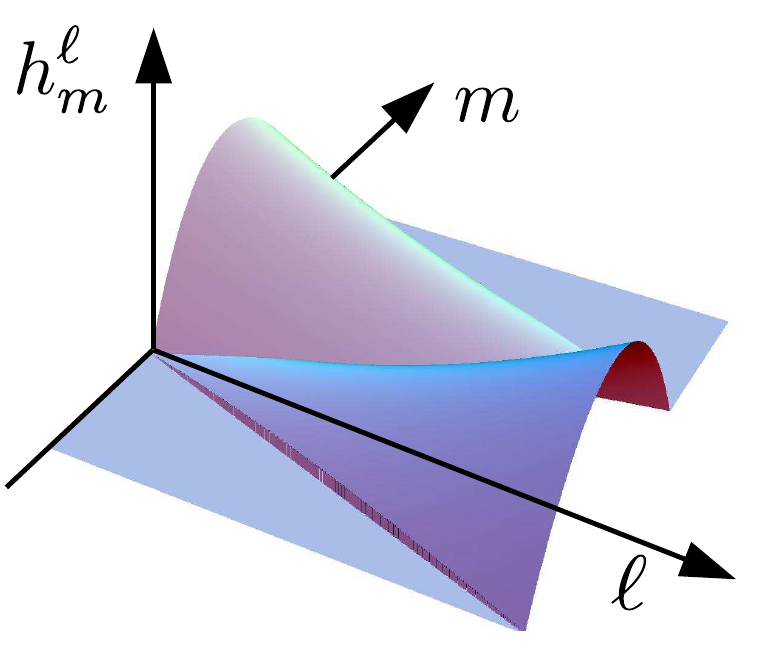}
\end{center}
\caption{Conformal dimension $h_m^\ell$ of parafermion fields.}
\label{hlm-plot}
\end{figure}

In the Read-Rezayi states, the parafermionic parts can change as follows:
the $m$ quantum number should stay fixed because it is related to the charge
by the $\Z_k$ parity rule, thus $\ell$ can change by an even integer:
the minimal value is $(\D \ell,\D m)=(\pm 2,0)$.
These transitions should reduce the value of the edge energy, i.e. of
$h_m^\ell$. The plot in Fig. \ref{hlm-plot} shows that the smallest values are
found on the diagonals $m=\pm \ell$ mod $2k$:
therefore, the peak patterns are analyzed starting from the low-lying
states $(\ell,m=\pm \ell)$.

\begin{figure}[t]
\begin{center}
\includegraphics[width=5cm]{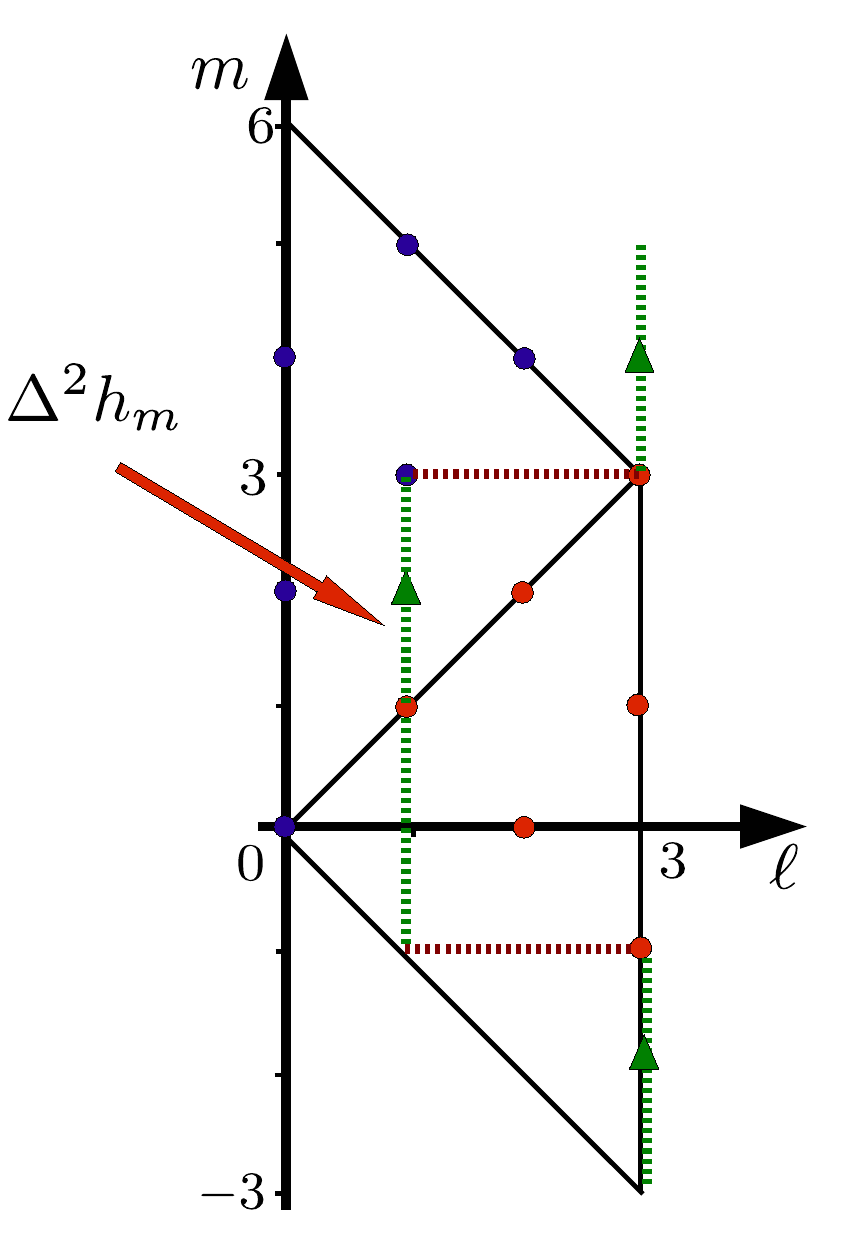}
\end{center}
\caption{Zig-zag path connecting the $\Z_3$ parafermion fields involved in
the evaluation of Coulomb peaks with relaxation: 
the read segments are relaxation transitions.}
\label{zig-zag}
\end{figure}

Let us consider for example the initial state $(\ell,m)=(1,1)$, 
as drawn in the parafermion diagram of Fig. \ref{zig-zag}. 
The first peak is found by comparing with the energy levels
of the next term in the same $\ell=1$ sector, i.e in $\theta^1_1$, which is 
$(1,3)$ (joined by a green line in Fig. \ref{zig-zag}); 
then the higher energy of the latter allows the relaxation,
$(1,3) \to (3,3)$, (red line), bringing to the $\ell=3$ sector. 
The next peak is therefore obtained by comparing $(3,3)$ to 
$(3,-1)$ , followed by the relaxation $(3,-1)\to (1,-1)$; the next
peak compares $(1,-1)$ with $(1,1)$ and so on.

Therefore, the peak patterns with relaxation are obtained by
comparing points in the $(\ell,m)$ plane that are reached
by zig-zag walking along the diagonals, that correspond to 
minimal edge energies (Fig. \ref{zig-zag}).
The relaxations are achieved by fusing the edge fields with the
$(\ell,m)=(2,0)$ parafermion that is always present in the spectrum
(higher $\D \ell$ transitions have larger energies and are never reached).
The peak distances $\D \s$ are computed as in section 4.2:
\be
\D \s=\s^{\ell\pm 2}_{\b+1}-\s^\ell_\b=\frac{\hat p}{k}+ 
\frac{v_n}{v}\left(
h^{\ell\pm 2}_{2\b+4}-h^{\ell\pm 2}_{2\b+2}
-h^\ell_{2\b+2}+h^\ell_{2\b} \right)\ ,
\label{pf-relax}
\ee
where the $\ell\pm 2$ point is along the zig-zag path.  These
derivatives are independent of $\ell$ and their values are already
known: the separations $\D \s$ acquire the extra contribution $v_n/v$
when the midpoint $m=2\b+2$ stays on a diagonal.

The resulting peak patterns are the following,
\be
\begin{array}{lcl}
k\ {\rm even,\ any}\ \ell:  
&\left(\frac{k}{2}\right)\ \left(\frac{k}{2}\right) &
 {\rm groups},
\\
k\ {\rm odd,\ any}\ \ \ell:  
&\left(\frac{k+1}{2}\right)\ \left(\frac{k-1}{2}\right) &
 {\rm groups}.
\end{array}
\label{pf-patt-rel}
\ee
They do not depend on the (starting) $\ell$ value;
in particular, the even-odd effect in the Pfaffian state doe not take
place in presence of bulk-edge relaxations. 
Form the experimental point of view,
relaxations could take place on long time scales, and thus be controlled,
but this is still unclear \cite{schou}.

Let us remark that
in the Jain hierarchical states of section 3, bulk-edge relaxations
corresponding to change of sectors $\theta_a$ (\ref{z-hier}) are not possible,
because any charge component, $K_\l$, appears only once in the spectrum
(cf.(\ref{z-ex})), so it is uniquely coupled to a neutral part.


\section{Discussion and conclusions}

In this paper, we described the CFT partition functions of quantum Hall
states: we introduced and solved the modular invariance conditions
of the annulus geometry, and obtained the disk expressions
describing isolated Hall droplets.
We stressed that the partition functions are defining quantities
of rational CFTs: they clearly display the sectors of the Hilbert space,
and account for the multiplicities of states and the fusion rules.

We used the disk partition functions to readily
analyze the degenerate edge states leading to Coulomb blockade
conductance peaks.  We found that the peak patterns can be non-trivial 
whenever the edge excitations involve neutral components;
generically, the modulation in the peak distances are of the
order $O(v_n/v)$, the ratio of velocities of neutral and charged
excitations.

The peak modulations can occur both in Abelian and non-Abelian Hall
states, but they can have different characteristics.
In the case of the Abelian Jain hierarchical states, the patterns
are independent of the presence of quasiparticles in the bulk and
bulk-edge relaxation phenomena are not possible.  Furthermore, the
multiplicities of edge excitations, due to the extended
symmetry of the multicomponent edge, could be observed by a careful analysis
of the peak sequences.  Two alternative theories for the
Jain states were also considered, that have no multiplicities.

The Coulomb peak patterns of non-Abelian Read-Rezayi states
do depend on the presence of bulk quasiparticles; however,
bulk-edge relaxation processes could be possible that would erase such 
dependence. We remark that relaxation phenomena are generically
possible in non-Abelian states.

Although the modulation of conductance peaks is not by itself 
a characteristic feature of non-Abelian statistics of excitations, 
the detailed features of the patterns can test the CFT description,
in particular the qualitative properties of its Hilbert space.
Isolated droplets, such as those involved in Coulomb blockade 
experiments, can provide rather clean experimental signals, that
may not reveal many features of Hall states but are nevertheless
very interesting.
In particular, the fine structure of conductance peaks in the Jain states
is a signature of the composite edge structure: from its lifting, 
one could check additional interactions in the Hamiltonian,  
inter-edge couplings and edge reconstruction effects etc. \cite{agam}. 
 
The chiral partition functions are useful in many contexts, besides the
study of Coulomb blockade presented in this paper.
In the future, we want to use them for further model building
of non-Abelian states and for the description of the
topological entanglement entropy of Hall droplets \cite{tee}.
The determination of $Z_{\rm disk}$ for other relevant Hall states,
such the Ardonne-Schoutens non-Abelian spin singlet states
\cite{ard}, will also presented in another publication.

While writing this report, we received the paper \cite{cb-last} where Coulomb
blockade patterns were also found for several Hall states.

\noindent{\bf Acknowledgments}

We thank L.S. Georgiev for collaboration in the early stages of this
project. I. D. Rodriguez, K. Schoutens, A. Stern and W. Kang are acknowledged
for interesting discussions.  A.C. would like to thank the hospitality of the
G. Galilei Institute for Theoretical Physics, Florence, and of Nordita,
Stockholm.  G.R.Z. acknowledges INFN and the Department of Physics, University
of Florence for partial support.  G.R.Z. is a fellow of CONICET.  This work
was partially supported by the ESF programme INSTANS and by a PRIN grant of
Italian Ministry of Education and Research.

\appendix


\section{Modular forms and functions}

In this appendix, we briefly review some of the properties of
modular invariant functions \cite{cz}.
Form section 2.2, we recall that annulus partition functions of 
quantum Hall states should contain fermionic excitations and be invariant
 under the $S$ and $T^2$ transformations.
They generate the subgroup $\Gamma_\theta \subset \Gamma$ which is
isomorphic to $\Gamma^0(2)=T\Gamma_\theta T^{-1}=\left\{
(a,b,c,d) \in \Gamma \vert b=0 \ {\rm mod}\ 2 \right\}$.
The transformations $ST^2S$ and $S$ generate the subgroup $\Gamma (2)$
of the modular group $\Gamma=SL(2,\Z)/\Z_2$ of transformations
(\ref{fracl}) which are subjected to the conditions $(a,d)$ odd and $(b,c)$
even.  

In the study of the functions on the torus, one naturally
encounters the modular forms $F(z)$, which transform under (\ref{fracl}) as:
\be
F\left( \frac{a\tau +b}{c\tau +d}\ \right) = \varepsilon\ 
\left(c\tau +d\right)^{\beta} F(\tau)\ ,
\label{modf}\ee
where $\varepsilon$ is a phase and $\beta$ is the weight of the modular form.
A modular function has weight $\beta=0$.
The simplest example of a modular form is the Dedekind function,
\be
\eta\left(\tau \right) =q^{\disp 1/24}\
\prod_{k=1}^\infty \left( 1-q^k \right)\
= q^{\disp 1/24}\ \sum_{n\in \Z} (-1)^n q^{\disp\ {1\over 2} n(3n+1)}\ ,
\qquad q={\rm e}^{\disp\ i2\pi \tau}\ .
\label{dede}\ee
The last equality is known as Euler's pentagonal
identity, and it is a consequence of Jacobi's triple product
identity \cite{cft}:
\be
\prod_{n=1}^{\infty} (1-q^n ) (1 + q^{n-1/2} w ) (1+q^{n-1/2}w^{-1})
= \sum_{n\in{\rm Z}} q^{n^2/2}  w^n \ ,
\label{jacobi}
\ee
after replacing in (\ref{jacobi}) $q\to q^3$ and $w\to -q^{-1/2}$.
Under the two generators $T:\ \tau\to\tau +1$ and $S:\ \tau\to -1/\tau$
of the modular group, the transformations laws of $\eta(\tau)$ are,
\ba
T:\ \eta(\tau +1) & = & {\rm e}^{\disp\ 2i\pi/24} \eta(\tau)\
,\label{tteta}\\
S:\ \eta(-1/\tau) & = & \left( -i\tau\right)^{1/2} \eta(\tau)\ .
\label{tseta}\ea
The proof of Eq.(\ref{tteta}) is straightforward, and that of
Eq.(\ref{tseta}) follows from the
application of Poisson's resummation formula,
\be
\sum_{n\in \Z} f(n)\ =\ \sum_{p\in \Z}\ \int_{-\infty}^{+\infty}
dx\ f(x)\ {\rm e}^{\disp\ 2i\pi px}\ ,
\label{poisson}\ee
to the r.h.s. of Eq.(\ref{dede}).
Under a general transformation (\ref{fracl}) we therefore have
\be
\eta\left(\frac{a\tau +b}{c\tau +d}\right) = \varepsilon_A
\left(c\tau +d\right)^{1/2} \eta(\tau)\ ,
\label{tetasl}
\ee
where $\varepsilon_A$ is a 24$th$ root of unity. Thus,
the Dedekind is a modular form of weight $1/2$.

Another important example of a modular form considered
in section two is the theta function with characteristics
$a$ and $b$, which is a map ${\cal F}\times{\C}\to{\C}$ 
defined by:
\be
\Theta\left[{ a \atop b}\right]
\left(\zeta\vert\ \tau\right) = \sum_{n\in \Z}\
{\rm e}^{\disp\ i\pi \tau (n+a)^2 + i2\pi (n+a)(\zeta + b) }\ .
\label{thetaab}\ee
In the case of Laughlin fluids, one has $b=0$ and $a=\lambda/p$,
with $\lambda =1,2,\dots,p$. 
The transformation properties of (\ref{thetaf}), Eq.(\ref{chitr})
follow easily from its definition. The only non-trivial
calculation regards the $S$ transformation, which can be done
following the example of the Dedekind function.
It is also easy to verify that (\ref{thetaab}) is a modular
form of weight $1/2$. It follows that the quotient
of the theta function (\ref{thetaab}) by the Dedekind function
(\ref{dede}) is a modular function.
In section 2 we introduced the notation:
\be
K\left(\t,\z;q \right) =\frac{1}{\eta}
\Th \left[{\l/q \atop 0 }\right] \left(\z \vert q\t \right) \ .
\ee


\end{document}